\documentclass[longauth]{aa}

\usepackage{graphicx}
\usepackage{natbib}
\usepackage{scalerel}

\usepackage[table]{xcolor}

\newcommand{\vect}[1]{\vec{#1}}
\newcommand{\Msun}{{\, M}_{\odot}}
\newcommand{\wslap}{\texttt{WSLAP+}\,}

\usepackage{txfonts}
\usepackage[pdfencoding=auto,psdextra]{hyperref}
\hypersetup{
    colorlinks=true,
    linkcolor=blue,
    filecolor=magenta,      
    urlcolor=blue,
    citecolor=blue
}
\makeatletter
\renewcommand*\aa@pageof{, page \thepage{} of \pageref*{LastPage}}
\makeatother

\usepackage[utf8]{inputenc}

\usepackage[switch, modulo]{lineno}

\usepackage{euclid} 
 
\begin{document}
%
%
   \title{\Euclid: Early Release Observations. A combined strong and weak lensing solution for Abell 2390 beyond its virial radius\thanks{This paper is published on behalf of the Euclid Consortium.}}   
%

\newcommand{\orcid}[1]{} 
\author{J.~M.~Diego\orcid{0000-0001-9065-3926}\thanks{\email{jdiego@ifca.unican.es}}\inst{\ref{aff1}}
\and G.~Congedo\orcid{0000-0003-2508-0046}\inst{\ref{aff2}}
\and R.~Gavazzi\orcid{0000-0002-5540-6935}\inst{\ref{aff3},\ref{aff4}}
\and T.~Schrabback\orcid{0000-0002-6987-7834}\inst{\ref{aff5}}
\and H.~Atek\orcid{0000-0002-7570-0824}\inst{\ref{aff4}}
\and B.~Jain\orcid{0000-0002-8220-3973}\inst{\ref{aff6}}
\and J.~R.~Weaver\orcid{0000-0003-1614-196X}\inst{\ref{aff7}}
\and Y.~Kang\orcid{0009-0000-8588-7250}\inst{\ref{aff8}}
\and W.~G.~Hartley\inst{\ref{aff8}}
\and G.~Mahler\orcid{0000-0003-3266-2001}\inst{\ref{aff9},\ref{aff10},\ref{aff11}}
\and N.~Okabe\orcid{0000-0003-2898-0728}\inst{\ref{aff12},\ref{aff13},\ref{aff14}}
\and J.~B.~Golden-Marx\orcid{0000-0002-6394-045X}\inst{\ref{aff15}}
\and M.~Meneghetti\orcid{0000-0003-1225-7084}\inst{\ref{aff16},\ref{aff17}}
\and J.~M.~Palencia\orcid{0000-0003-0942-817X}\inst{\ref{aff1}}
\and M.~Kluge\orcid{0000-0002-9618-2552}\inst{\ref{aff18}}
\and R.~Laureijs\inst{\ref{aff19}}
\and T.~Saifollahi\orcid{0000-0002-9554-7660}\inst{\ref{aff20}}
\and M.~Schirmer\orcid{0000-0003-2568-9994}\inst{\ref{aff21}}
\and C.~Stone\orcid{0000-0002-9086-6398}\inst{\ref{aff22},\ref{aff23},\ref{aff24}}
\and M.~Jauzac\orcid{0000-0003-1974-8732}\inst{\ref{aff10},\ref{aff11},\ref{aff25},\ref{aff26}}
\and D.~Scott\orcid{0000-0002-6878-9840}\inst{\ref{aff27}}
\and B.~Altieri\orcid{0000-0003-3936-0284}\inst{\ref{aff28}}
\and A.~Amara\inst{\ref{aff29}}
\and S.~Andreon\orcid{0000-0002-2041-8784}\inst{\ref{aff30}}
\and N.~Auricchio\orcid{0000-0003-4444-8651}\inst{\ref{aff16}}
\and C.~Baccigalupi\orcid{0000-0002-8211-1630}\inst{\ref{aff31},\ref{aff32},\ref{aff33},\ref{aff34}}
\and M.~Baldi\orcid{0000-0003-4145-1943}\inst{\ref{aff35},\ref{aff16},\ref{aff17}}
\and S.~Bardelli\orcid{0000-0002-8900-0298}\inst{\ref{aff16}}
\and P.~Battaglia\orcid{0000-0002-7337-5909}\inst{\ref{aff16}}
\and A.~Biviano\orcid{0000-0002-0857-0732}\inst{\ref{aff32},\ref{aff31}}
\and E.~Branchini\orcid{0000-0002-0808-6908}\inst{\ref{aff36},\ref{aff37},\ref{aff30}}
\and M.~Brescia\orcid{0000-0001-9506-5680}\inst{\ref{aff38},\ref{aff39}}
\and J.~Brinchmann\orcid{0000-0003-4359-8797}\inst{\ref{aff40},\ref{aff41},\ref{aff42}}
\and S.~Camera\orcid{0000-0003-3399-3574}\inst{\ref{aff43},\ref{aff44},\ref{aff45}}
\and G.~Ca\~nas-Herrera\orcid{0000-0003-2796-2149}\inst{\ref{aff46},\ref{aff47},\ref{aff48}}
\and G.~P.~Candini\orcid{0000-0001-9481-8206}\inst{\ref{aff49}}
\and V.~Capobianco\orcid{0000-0002-3309-7692}\inst{\ref{aff45}}
\and C.~Carbone\orcid{0000-0003-0125-3563}\inst{\ref{aff50}}
\and V.~F.~Cardone\inst{\ref{aff51},\ref{aff52}}
\and J.~Carretero\orcid{0000-0002-3130-0204}\inst{\ref{aff53},\ref{aff54}}
\and S.~Casas\orcid{0000-0002-4751-5138}\inst{\ref{aff55}}
\and M.~Castellano\orcid{0000-0001-9875-8263}\inst{\ref{aff51}}
\and G.~Castignani\orcid{0000-0001-6831-0687}\inst{\ref{aff16}}
\and S.~Cavuoti\orcid{0000-0002-3787-4196}\inst{\ref{aff39},\ref{aff56}}
\and K.~C.~Chambers\orcid{0000-0001-6965-7789}\inst{\ref{aff57}}
\and A.~Cimatti\inst{\ref{aff58}}
\and C.~Colodro-Conde\inst{\ref{aff59}}
\and C.~J.~Conselice\orcid{0000-0003-1949-7638}\inst{\ref{aff60}}
\and L.~Conversi\orcid{0000-0002-6710-8476}\inst{\ref{aff61},\ref{aff28}}
\and Y.~Copin\orcid{0000-0002-5317-7518}\inst{\ref{aff62}}
\and F.~Courbin\orcid{0000-0003-0758-6510}\inst{\ref{aff63},\ref{aff64}}
\and H.~M.~Courtois\orcid{0000-0003-0509-1776}\inst{\ref{aff65}}
\and M.~Cropper\orcid{0000-0003-4571-9468}\inst{\ref{aff49}}
\and J.-C.~Cuillandre\orcid{0000-0002-3263-8645}\inst{\ref{aff66}}
\and A.~Da~Silva\orcid{0000-0002-6385-1609}\inst{\ref{aff67},\ref{aff68}}
\and H.~Degaudenzi\orcid{0000-0002-5887-6799}\inst{\ref{aff8}}
\and G.~De~Lucia\orcid{0000-0002-6220-9104}\inst{\ref{aff32}}
\and H.~Dole\orcid{0000-0002-9767-3839}\inst{\ref{aff69}}
\and M.~Douspis\orcid{0000-0003-4203-3954}\inst{\ref{aff69}}
\and F.~Dubath\orcid{0000-0002-6533-2810}\inst{\ref{aff8}}
\and X.~Dupac\inst{\ref{aff28}}
\and S.~Dusini\orcid{0000-0002-1128-0664}\inst{\ref{aff70}}
\and S.~Escoffier\orcid{0000-0002-2847-7498}\inst{\ref{aff71}}
\and M.~Farina\orcid{0000-0002-3089-7846}\inst{\ref{aff72}}
\and S.~Farrens\orcid{0000-0002-9594-9387}\inst{\ref{aff66}}
\and F.~Faustini\orcid{0000-0001-6274-5145}\inst{\ref{aff51},\ref{aff73}}
\and S.~Ferriol\inst{\ref{aff62}}
\and F.~Finelli\orcid{0000-0002-6694-3269}\inst{\ref{aff16},\ref{aff74}}
\and P.~Fosalba\orcid{0000-0002-1510-5214}\inst{\ref{aff75},\ref{aff76}}
\and N.~Fourmanoit\orcid{0009-0005-6816-6925}\inst{\ref{aff71}}
\and M.~Frailis\orcid{0000-0002-7400-2135}\inst{\ref{aff32}}
\and E.~Franceschi\orcid{0000-0002-0585-6591}\inst{\ref{aff16}}
\and M.~Fumana\orcid{0000-0001-6787-5950}\inst{\ref{aff50}}
\and S.~Galeotta\orcid{0000-0002-3748-5115}\inst{\ref{aff32}}
\and K.~George\orcid{0000-0002-1734-8455}\inst{\ref{aff77}}
\and B.~Gillis\orcid{0000-0002-4478-1270}\inst{\ref{aff2}}
\and C.~Giocoli\orcid{0000-0002-9590-7961}\inst{\ref{aff16},\ref{aff17}}
\and J.~Gracia-Carpio\inst{\ref{aff18}}
\and A.~Grazian\orcid{0000-0002-5688-0663}\inst{\ref{aff78}}
\and F.~Grupp\inst{\ref{aff18},\ref{aff77}}
\and L.~Guzzo\orcid{0000-0001-8264-5192}\inst{\ref{aff79},\ref{aff30},\ref{aff80}}
\and S.~V.~H.~Haugan\orcid{0000-0001-9648-7260}\inst{\ref{aff81}}
\and J.~Hoar\inst{\ref{aff28}}
\and W.~Holmes\inst{\ref{aff82}}
\and I.~M.~Hook\orcid{0000-0002-2960-978X}\inst{\ref{aff83}}
\and F.~Hormuth\inst{\ref{aff84}}
\and A.~Hornstrup\orcid{0000-0002-3363-0936}\inst{\ref{aff85},\ref{aff86}}
\and P.~Hudelot\inst{\ref{aff4}}
\and K.~Jahnke\orcid{0000-0003-3804-2137}\inst{\ref{aff21}}
\and M.~Jhabvala\inst{\ref{aff87}}
\and B.~Joachimi\orcid{0000-0001-7494-1303}\inst{\ref{aff88}}
\and E.~Keih\"anen\orcid{0000-0003-1804-7715}\inst{\ref{aff89}}
\and S.~Kermiche\orcid{0000-0002-0302-5735}\inst{\ref{aff71}}
\and M.~Kilbinger\orcid{0000-0001-9513-7138}\inst{\ref{aff66}}
\and B.~Kubik\orcid{0009-0006-5823-4880}\inst{\ref{aff62}}
\and K.~Kuijken\orcid{0000-0002-3827-0175}\inst{\ref{aff48}}
\and M.~K\"ummel\orcid{0000-0003-2791-2117}\inst{\ref{aff77}}
\and M.~Kunz\orcid{0000-0002-3052-7394}\inst{\ref{aff90}}
\and H.~Kurki-Suonio\orcid{0000-0002-4618-3063}\inst{\ref{aff91},\ref{aff92}}
\and A.~M.~C.~Le~Brun\orcid{0000-0002-0936-4594}\inst{\ref{aff93}}
\and D.~Le~Mignant\orcid{0000-0002-5339-5515}\inst{\ref{aff3}}
\and S.~Ligori\orcid{0000-0003-4172-4606}\inst{\ref{aff45}}
\and P.~B.~Lilje\orcid{0000-0003-4324-7794}\inst{\ref{aff81}}
\and V.~Lindholm\orcid{0000-0003-2317-5471}\inst{\ref{aff91},\ref{aff92}}
\and I.~Lloro\orcid{0000-0001-5966-1434}\inst{\ref{aff94}}
\and G.~Mainetti\orcid{0000-0003-2384-2377}\inst{\ref{aff95}}
\and D.~Maino\inst{\ref{aff79},\ref{aff50},\ref{aff80}}
\and E.~Maiorano\orcid{0000-0003-2593-4355}\inst{\ref{aff16}}
\and O.~Mansutti\orcid{0000-0001-5758-4658}\inst{\ref{aff32}}
\and O.~Marggraf\orcid{0000-0001-7242-3852}\inst{\ref{aff96}}
\and M.~Martinelli\orcid{0000-0002-6943-7732}\inst{\ref{aff51},\ref{aff52}}
\and N.~Martinet\orcid{0000-0003-2786-7790}\inst{\ref{aff3}}
\and F.~Marulli\orcid{0000-0002-8850-0303}\inst{\ref{aff97},\ref{aff16},\ref{aff17}}
\and R.~J.~Massey\orcid{0000-0002-6085-3780}\inst{\ref{aff11}}
\and E.~Medinaceli\orcid{0000-0002-4040-7783}\inst{\ref{aff16}}
\and S.~Mei\orcid{0000-0002-2849-559X}\inst{\ref{aff98},\ref{aff99}}
\and M.~Melchior\inst{\ref{aff100}}
\and Y.~Mellier\inst{\ref{aff101},\ref{aff4}}
\and E.~Merlin\orcid{0000-0001-6870-8900}\inst{\ref{aff51}}
\and G.~Meylan\inst{\ref{aff102}}
\and J.~J.~Mohr\orcid{0000-0002-6875-2087}\inst{\ref{aff103}}
\and A.~Mora\orcid{0000-0002-1922-8529}\inst{\ref{aff104}}
\and M.~Moresco\orcid{0000-0002-7616-7136}\inst{\ref{aff97},\ref{aff16}}
\and L.~Moscardini\orcid{0000-0002-3473-6716}\inst{\ref{aff97},\ref{aff16},\ref{aff17}}
\and E.~Munari\orcid{0000-0002-1751-5946}\inst{\ref{aff32},\ref{aff31}}
\and R.~Nakajima\orcid{0009-0009-1213-7040}\inst{\ref{aff96}}
\and C.~Neissner\orcid{0000-0001-8524-4968}\inst{\ref{aff105},\ref{aff54}}
\and R.~C.~Nichol\orcid{0000-0003-0939-6518}\inst{\ref{aff29}}
\and S.-M.~Niemi\orcid{0009-0005-0247-0086}\inst{\ref{aff46}}
\and C.~Padilla\orcid{0000-0001-7951-0166}\inst{\ref{aff105}}
\and S.~Paltani\orcid{0000-0002-8108-9179}\inst{\ref{aff8}}
\and F.~Pasian\orcid{0000-0002-4869-3227}\inst{\ref{aff32}}
\and K.~Pedersen\inst{\ref{aff106}}
\and W.~J.~Percival\orcid{0000-0002-0644-5727}\inst{\ref{aff107},\ref{aff108},\ref{aff109}}
\and V.~Pettorino\inst{\ref{aff46}}
\and S.~Pires\orcid{0000-0002-0249-2104}\inst{\ref{aff66}}
\and G.~Polenta\orcid{0000-0003-4067-9196}\inst{\ref{aff73}}
\and M.~Poncet\inst{\ref{aff110}}
\and L.~A.~Popa\inst{\ref{aff111}}
\and L.~Pozzetti\orcid{0000-0001-7085-0412}\inst{\ref{aff16}}
\and F.~Raison\orcid{0000-0002-7819-6918}\inst{\ref{aff18}}
\and A.~Renzi\orcid{0000-0001-9856-1970}\inst{\ref{aff112},\ref{aff70}}
\and J.~Rhodes\orcid{0000-0002-4485-8549}\inst{\ref{aff82}}
\and G.~Riccio\inst{\ref{aff39}}
\and E.~Romelli\orcid{0000-0003-3069-9222}\inst{\ref{aff32}}
\and M.~Roncarelli\orcid{0000-0001-9587-7822}\inst{\ref{aff16}}
\and R.~Saglia\orcid{0000-0003-0378-7032}\inst{\ref{aff77},\ref{aff18}}
\and Z.~Sakr\orcid{0000-0002-4823-3757}\inst{\ref{aff113},\ref{aff114},\ref{aff115}}
\and D.~Sapone\orcid{0000-0001-7089-4503}\inst{\ref{aff116}}
\and B.~Sartoris\orcid{0000-0003-1337-5269}\inst{\ref{aff77},\ref{aff32}}
\and P.~Schneider\orcid{0000-0001-8561-2679}\inst{\ref{aff96}}
\and A.~Secroun\orcid{0000-0003-0505-3710}\inst{\ref{aff71}}
\and G.~Seidel\orcid{0000-0003-2907-353X}\inst{\ref{aff21}}
\and M.~Seiffert\orcid{0000-0002-7536-9393}\inst{\ref{aff82}}
\and S.~Serrano\orcid{0000-0002-0211-2861}\inst{\ref{aff75},\ref{aff117},\ref{aff76}}
\and P.~Simon\inst{\ref{aff96}}
\and C.~Sirignano\orcid{0000-0002-0995-7146}\inst{\ref{aff112},\ref{aff70}}
\and G.~Sirri\orcid{0000-0003-2626-2853}\inst{\ref{aff17}}
\and L.~Stanco\orcid{0000-0002-9706-5104}\inst{\ref{aff70}}
\and J.~Steinwagner\orcid{0000-0001-7443-1047}\inst{\ref{aff18}}
\and P.~Tallada-Cresp\'{i}\orcid{0000-0002-1336-8328}\inst{\ref{aff53},\ref{aff54}}
\and A.~N.~Taylor\inst{\ref{aff2}}
\and I.~Tereno\orcid{0000-0002-4537-6218}\inst{\ref{aff67},\ref{aff118}}
\and N.~Tessore\orcid{0000-0002-9696-7931}\inst{\ref{aff88}}
\and S.~Toft\orcid{0000-0003-3631-7176}\inst{\ref{aff119},\ref{aff120}}
\and R.~Toledo-Moreo\orcid{0000-0002-2997-4859}\inst{\ref{aff121}}
\and F.~Torradeflot\orcid{0000-0003-1160-1517}\inst{\ref{aff54},\ref{aff53}}
\and I.~Tutusaus\orcid{0000-0002-3199-0399}\inst{\ref{aff114}}
\and L.~Valenziano\orcid{0000-0002-1170-0104}\inst{\ref{aff16},\ref{aff74}}
\and J.~Valiviita\orcid{0000-0001-6225-3693}\inst{\ref{aff91},\ref{aff92}}
\and T.~Vassallo\orcid{0000-0001-6512-6358}\inst{\ref{aff77},\ref{aff32}}
\and G.~Verdoes~Kleijn\orcid{0000-0001-5803-2580}\inst{\ref{aff19}}
\and A.~Veropalumbo\orcid{0000-0003-2387-1194}\inst{\ref{aff30},\ref{aff37},\ref{aff36}}
\and Y.~Wang\orcid{0000-0002-4749-2984}\inst{\ref{aff122}}
\and J.~Weller\orcid{0000-0002-8282-2010}\inst{\ref{aff77},\ref{aff18}}
\and G.~Zamorani\orcid{0000-0002-2318-301X}\inst{\ref{aff16}}
\and F.~M.~Zerbi\inst{\ref{aff30}}
\and E.~Zucca\orcid{0000-0002-5845-8132}\inst{\ref{aff16}}
\and M.~Bolzonella\orcid{0000-0003-3278-4607}\inst{\ref{aff16}}
\and C.~Burigana\orcid{0000-0002-3005-5796}\inst{\ref{aff123},\ref{aff74}}
\and L.~Gabarra\orcid{0000-0002-8486-8856}\inst{\ref{aff124}}
\and J.~Mart\'{i}n-Fleitas\orcid{0000-0002-8594-569X}\inst{\ref{aff125}}
\and S.~Matthew\orcid{0000-0001-8448-1697}\inst{\ref{aff2}}
\and V.~Scottez\orcid{0009-0008-3864-940X}\inst{\ref{aff101},\ref{aff126}}
\and M.~Sereno\orcid{0000-0003-0302-0325}\inst{\ref{aff16},\ref{aff17}}
\and M.~Viel\orcid{0000-0002-2642-5707}\inst{\ref{aff31},\ref{aff32},\ref{aff34},\ref{aff33},\ref{aff127}}}
										   
\institute{Instituto de F\'isica de Cantabria, Edificio Juan Jord\'a, Avenida de los Castros, 39005 Santander, Spain\label{aff1}
\and
Institute for Astronomy, University of Edinburgh, Royal Observatory, Blackford Hill, Edinburgh EH9 3HJ, UK\label{aff2}
\and
Aix-Marseille Universit\'e, CNRS, CNES, LAM, Marseille, France\label{aff3}
\and
Institut d'Astrophysique de Paris, UMR 7095, CNRS, and Sorbonne Universit\'e, 98 bis boulevard Arago, 75014 Paris, France\label{aff4}
\and
Universit\"at Innsbruck, Institut f\"ur Astro- und Teilchenphysik, Technikerstr. 25/8, 6020 Innsbruck, Austria\label{aff5}
\and
Department of Physics and Astronomy, University of Pennsylvania, Philadelphia, PA 19146, USA\label{aff6}
\and
Department of Astronomy, University of Massachusetts, Amherst, MA 01003, USA\label{aff7}
\and
Department of Astronomy, University of Geneva, ch. d'Ecogia 16, 1290 Versoix, Switzerland\label{aff8}
\and
STAR Institute, University of Li{\`e}ge, Quartier Agora, All\'ee du six Ao\^ut 19c, 4000 Li\`ege, Belgium\label{aff9}
\and
Department of Physics, Centre for Extragalactic Astronomy, Durham University, South Road, Durham, DH1 3LE, UK\label{aff10}
\and
Department of Physics, Institute for Computational Cosmology, Durham University, South Road, Durham, DH1 3LE, UK\label{aff11}
\and
Physics Program, Graduate School of Advanced Science and Engineering, Hiroshima University, 1-3-1 Kagamiyama, Higashi-Hiroshima, Hiroshima 739-8526, Japan\label{aff12}
\and
Hiroshima Astrophysical Science Center, Hiroshima University, 1-3-1 Kagamiyama, Higashi-Hiroshima, Hiroshima 739-8526, Japan\label{aff13}
\and
Core Research for Energetic Universe, Hiroshima University, 1-3-1, Kagamiyama, Higashi-Hiroshima, Hiroshima 739-8526, Japan\label{aff14}
\and
School of Physics and Astronomy, University of Nottingham, University Park, Nottingham NG7 2RD, UK\label{aff15}
\and
INAF-Osservatorio di Astrofisica e Scienza dello Spazio di Bologna, Via Piero Gobetti 93/3, 40129 Bologna, Italy\label{aff16}
\and
INFN-Sezione di Bologna, Viale Berti Pichat 6/2, 40127 Bologna, Italy\label{aff17}
\and
Max Planck Institute for Extraterrestrial Physics, Giessenbachstr. 1, 85748 Garching, Germany\label{aff18}
\and
Kapteyn Astronomical Institute, University of Groningen, PO Box 800, 9700 AV Groningen, The Netherlands\label{aff19}
\and
Universit\'e de Strasbourg, CNRS, Observatoire astronomique de Strasbourg, UMR 7550, 67000 Strasbourg, France\label{aff20}
\and
Max-Planck-Institut f\"ur Astronomie, K\"onigstuhl 17, 69117 Heidelberg, Germany\label{aff21}
\and
Department of Physics, Universit\'{e} de Montr\'{e}al, 2900 Edouard Montpetit Blvd, Montr\'{e}al, Qu\'{e}bec H3T 1J4, Canada\label{aff22}
\and
Ciela Institute - Montr{\'e}al Institute for Astrophysical Data Analysis and Machine Learning, Montr{\'e}al, Qu{\'e}bec, Canada\label{aff23}
\and
Mila - Qu{\'e}bec Artificial Intelligence Institute, Montr{\'e}al, Qu{\'e}bec, Canada\label{aff24}
\and
Astrophysics Research Centre, University of KwaZulu-Natal, Westville Campus, Durban 4041, South Africa\label{aff25}
\and
School of Mathematics, Statistics \& Computer Science, University of KwaZulu-Natal, Westville Campus, Durban 4041, South Africa\label{aff26}
\and
Department of Physics and Astronomy, University of British Columbia, Vancouver, BC V6T 1Z1, Canada\label{aff27}
\and
ESAC/ESA, Camino Bajo del Castillo, s/n., Urb. Villafranca del Castillo, 28692 Villanueva de la Ca\~nada, Madrid, Spain\label{aff28}
\and
School of Mathematics and Physics, University of Surrey, Guildford, Surrey, GU2 7XH, UK\label{aff29}
\and
INAF-Osservatorio Astronomico di Brera, Via Brera 28, 20122 Milano, Italy\label{aff30}
\and
IFPU, Institute for Fundamental Physics of the Universe, via Beirut 2, 34151 Trieste, Italy\label{aff31}
\and
INAF-Osservatorio Astronomico di Trieste, Via G. B. Tiepolo 11, 34143 Trieste, Italy\label{aff32}
\and
INFN, Sezione di Trieste, Via Valerio 2, 34127 Trieste TS, Italy\label{aff33}
\and
SISSA, International School for Advanced Studies, Via Bonomea 265, 34136 Trieste TS, Italy\label{aff34}
\and
Dipartimento di Fisica e Astronomia, Universit\`a di Bologna, Via Gobetti 93/2, 40129 Bologna, Italy\label{aff35}
\and
Dipartimento di Fisica, Universit\`a di Genova, Via Dodecaneso 33, 16146, Genova, Italy\label{aff36}
\and
INFN-Sezione di Genova, Via Dodecaneso 33, 16146, Genova, Italy\label{aff37}
\and
Department of Physics "E. Pancini", University Federico II, Via Cinthia 6, 80126, Napoli, Italy\label{aff38}
\and
INAF-Osservatorio Astronomico di Capodimonte, Via Moiariello 16, 80131 Napoli, Italy\label{aff39}
\and
Instituto de Astrof\'isica e Ci\^encias do Espa\c{c}o, Universidade do Porto, CAUP, Rua das Estrelas, PT4150-762 Porto, Portugal\label{aff40}
\and
Faculdade de Ci\^encias da Universidade do Porto, Rua do Campo de Alegre, 4150-007 Porto, Portugal\label{aff41}
\and
European Southern Observatory, Karl-Schwarzschild-Str.~2, 85748 Garching, Germany\label{aff42}
\and
Dipartimento di Fisica, Universit\`a degli Studi di Torino, Via P. Giuria 1, 10125 Torino, Italy\label{aff43}
\and
INFN-Sezione di Torino, Via P. Giuria 1, 10125 Torino, Italy\label{aff44}
\and
INAF-Osservatorio Astrofisico di Torino, Via Osservatorio 20, 10025 Pino Torinese (TO), Italy\label{aff45}
\and
European Space Agency/ESTEC, Keplerlaan 1, 2201 AZ Noordwijk, The Netherlands\label{aff46}
\and
Institute Lorentz, Leiden University, Niels Bohrweg 2, 2333 CA Leiden, The Netherlands\label{aff47}
\and
Leiden Observatory, Leiden University, Einsteinweg 55, 2333 CC Leiden, The Netherlands\label{aff48}
\and
Mullard Space Science Laboratory, University College London, Holmbury St Mary, Dorking, Surrey RH5 6NT, UK\label{aff49}
\and
INAF-IASF Milano, Via Alfonso Corti 12, 20133 Milano, Italy\label{aff50}
\and
INAF-Osservatorio Astronomico di Roma, Via Frascati 33, 00078 Monteporzio Catone, Italy\label{aff51}
\and
INFN-Sezione di Roma, Piazzale Aldo Moro, 2 - c/o Dipartimento di Fisica, Edificio G. Marconi, 00185 Roma, Italy\label{aff52}
\and
Centro de Investigaciones Energ\'eticas, Medioambientales y Tecnol\'ogicas (CIEMAT), Avenida Complutense 40, 28040 Madrid, Spain\label{aff53}
\and
Port d'Informaci\'{o} Cient\'{i}fica, Campus UAB, C. Albareda s/n, 08193 Bellaterra (Barcelona), Spain\label{aff54}
\and
Institute for Theoretical Particle Physics and Cosmology (TTK), RWTH Aachen University, 52056 Aachen, Germany\label{aff55}
\and
INFN section of Naples, Via Cinthia 6, 80126, Napoli, Italy\label{aff56}
\and
Institute for Astronomy, University of Hawaii, 2680 Woodlawn Drive, Honolulu, HI 96822, USA\label{aff57}
\and
Dipartimento di Fisica e Astronomia "Augusto Righi" - Alma Mater Studiorum Universit\`a di Bologna, Viale Berti Pichat 6/2, 40127 Bologna, Italy\label{aff58}
\and
Instituto de Astrof\'{\i}sica de Canarias, V\'{\i}a L\'actea, 38205 La Laguna, Tenerife, Spain\label{aff59}
\and
Jodrell Bank Centre for Astrophysics, Department of Physics and Astronomy, University of Manchester, Oxford Road, Manchester M13 9PL, UK\label{aff60}
\and
European Space Agency/ESRIN, Largo Galileo Galilei 1, 00044 Frascati, Roma, Italy\label{aff61}
\and
Universit\'e Claude Bernard Lyon 1, CNRS/IN2P3, IP2I Lyon, UMR 5822, Villeurbanne, F-69100, France\label{aff62}
\and
Institut de Ci\`{e}ncies del Cosmos (ICCUB), Universitat de Barcelona (IEEC-UB), Mart\'{i} i Franqu\`{e}s 1, 08028 Barcelona, Spain\label{aff63}
\and
Instituci\'o Catalana de Recerca i Estudis Avan\c{c}ats (ICREA), Passeig de Llu\'{\i}s Companys 23, 08010 Barcelona, Spain\label{aff64}
\and
UCB Lyon 1, CNRS/IN2P3, IUF, IP2I Lyon, 4 rue Enrico Fermi, 69622 Villeurbanne, France\label{aff65}
\and
Universit\'e Paris-Saclay, Universit\'e Paris Cit\'e, CEA, CNRS, AIM, 91191, Gif-sur-Yvette, France\label{aff66}
\and
Departamento de F\'isica, Faculdade de Ci\^encias, Universidade de Lisboa, Edif\'icio C8, Campo Grande, PT1749-016 Lisboa, Portugal\label{aff67}
\and
Instituto de Astrof\'isica e Ci\^encias do Espa\c{c}o, Faculdade de Ci\^encias, Universidade de Lisboa, Campo Grande, 1749-016 Lisboa, Portugal\label{aff68}
\and
Universit\'e Paris-Saclay, CNRS, Institut d'astrophysique spatiale, 91405, Orsay, France\label{aff69}
\and
INFN-Padova, Via Marzolo 8, 35131 Padova, Italy\label{aff70}
\and
Aix-Marseille Universit\'e, CNRS/IN2P3, CPPM, Marseille, France\label{aff71}
\and
INAF-Istituto di Astrofisica e Planetologia Spaziali, via del Fosso del Cavaliere, 100, 00100 Roma, Italy\label{aff72}
\and
Space Science Data Center, Italian Space Agency, via del Politecnico snc, 00133 Roma, Italy\label{aff73}
\and
INFN-Bologna, Via Irnerio 46, 40126 Bologna, Italy\label{aff74}
\and
Institut d'Estudis Espacials de Catalunya (IEEC),  Edifici RDIT, Campus UPC, 08860 Castelldefels, Barcelona, Spain\label{aff75}
\and
Institute of Space Sciences (ICE, CSIC), Campus UAB, Carrer de Can Magrans, s/n, 08193 Barcelona, Spain\label{aff76}
\and
Universit\"ats-Sternwarte M\"unchen, Fakult\"at f\"ur Physik, Ludwig-Maximilians-Universit\"at M\"unchen, Scheinerstrasse 1, 81679 M\"unchen, Germany\label{aff77}
\and
INAF-Osservatorio Astronomico di Padova, Via dell'Osservatorio 5, 35122 Padova, Italy\label{aff78}
\and
Dipartimento di Fisica "Aldo Pontremoli", Universit\`a degli Studi di Milano, Via Celoria 16, 20133 Milano, Italy\label{aff79}
\and
INFN-Sezione di Milano, Via Celoria 16, 20133 Milano, Italy\label{aff80}
\and
Institute of Theoretical Astrophysics, University of Oslo, P.O. Box 1029 Blindern, 0315 Oslo, Norway\label{aff81}
\and
Jet Propulsion Laboratory, California Institute of Technology, 4800 Oak Grove Drive, Pasadena, CA, 91109, USA\label{aff82}
\and
Department of Physics, Lancaster University, Lancaster, LA1 4YB, UK\label{aff83}
\and
Felix Hormuth Engineering, Goethestr. 17, 69181 Leimen, Germany\label{aff84}
\and
Technical University of Denmark, Elektrovej 327, 2800 Kgs. Lyngby, Denmark\label{aff85}
\and
Cosmic Dawn Center (DAWN), Denmark\label{aff86}
\and
NASA Goddard Space Flight Center, Greenbelt, MD 20771, USA\label{aff87}
\and
Department of Physics and Astronomy, University College London, Gower Street, London WC1E 6BT, UK\label{aff88}
\and
Department of Physics and Helsinki Institute of Physics, Gustaf H\"allstr\"omin katu 2, 00014 University of Helsinki, Finland\label{aff89}
\and
Universit\'e de Gen\`eve, D\'epartement de Physique Th\'eorique and Centre for Astroparticle Physics, 24 quai Ernest-Ansermet, CH-1211 Gen\`eve 4, Switzerland\label{aff90}
\and
Department of Physics, P.O. Box 64, 00014 University of Helsinki, Finland\label{aff91}
\and
Helsinki Institute of Physics, Gustaf H{\"a}llstr{\"o}min katu 2, University of Helsinki, Helsinki, Finland\label{aff92}
\and
Laboratoire d'etude de l'Univers et des phenomenes eXtremes, Observatoire de Paris, Universit\'e PSL, Sorbonne Universit\'e, CNRS, 92190 Meudon, France\label{aff93}
\and
SKA Observatory, Jodrell Bank, Lower Withington, Macclesfield, Cheshire SK11 9FT, UK\label{aff94}
\and
Centre de Calcul de l'IN2P3/CNRS, 21 avenue Pierre de Coubertin 69627 Villeurbanne Cedex, France\label{aff95}
\and
Universit\"at Bonn, Argelander-Institut f\"ur Astronomie, Auf dem H\"ugel 71, 53121 Bonn, Germany\label{aff96}
\and
Dipartimento di Fisica e Astronomia "Augusto Righi" - Alma Mater Studiorum Universit\`a di Bologna, via Piero Gobetti 93/2, 40129 Bologna, Italy\label{aff97}
\and
Universit\'e Paris Cit\'e, CNRS, Astroparticule et Cosmologie, 75013 Paris, France\label{aff98}
\and
CNRS-UCB International Research Laboratory, Centre Pierre Bin\'etruy, IRL2007, CPB-IN2P3, Berkeley, USA\label{aff99}
\and
University of Applied Sciences and Arts of Northwestern Switzerland, School of Engineering, 5210 Windisch, Switzerland\label{aff100}
\and
Institut d'Astrophysique de Paris, 98bis Boulevard Arago, 75014, Paris, France\label{aff101}
\and
Institute of Physics, Laboratory of Astrophysics, Ecole Polytechnique F\'ed\'erale de Lausanne (EPFL), Observatoire de Sauverny, 1290 Versoix, Switzerland\label{aff102}
\and
University Observatory, LMU Faculty of Physics, Scheinerstrasse 1, 81679 Munich, Germany\label{aff103}
\and
Telespazio UK S.L. for European Space Agency (ESA), Camino bajo del Castillo, s/n, Urbanizacion Villafranca del Castillo, Villanueva de la Ca\~nada, 28692 Madrid, Spain\label{aff104}
\and
Institut de F\'{i}sica d'Altes Energies (IFAE), The Barcelona Institute of Science and Technology, Campus UAB, 08193 Bellaterra (Barcelona), Spain\label{aff105}
\and
DARK, Niels Bohr Institute, University of Copenhagen, Jagtvej 155, 2200 Copenhagen, Denmark\label{aff106}
\and
Waterloo Centre for Astrophysics, University of Waterloo, Waterloo, Ontario N2L 3G1, Canada\label{aff107}
\and
Department of Physics and Astronomy, University of Waterloo, Waterloo, Ontario N2L 3G1, Canada\label{aff108}
\and
Perimeter Institute for Theoretical Physics, Waterloo, Ontario N2L 2Y5, Canada\label{aff109}
\and
Centre National d'Etudes Spatiales -- Centre spatial de Toulouse, 18 avenue Edouard Belin, 31401 Toulouse Cedex 9, France\label{aff110}
\and
Institute of Space Science, Str. Atomistilor, nr. 409 M\u{a}gurele, Ilfov, 077125, Romania\label{aff111}
\and
Dipartimento di Fisica e Astronomia "G. Galilei", Universit\`a di Padova, Via Marzolo 8, 35131 Padova, Italy\label{aff112}
\and
Institut f\"ur Theoretische Physik, University of Heidelberg, Philosophenweg 16, 69120 Heidelberg, Germany\label{aff113}
\and
Institut de Recherche en Astrophysique et Plan\'etologie (IRAP), Universit\'e de Toulouse, CNRS, UPS, CNES, 14 Av. Edouard Belin, 31400 Toulouse, France\label{aff114}
\and
Universit\'e St Joseph; Faculty of Sciences, Beirut, Lebanon\label{aff115}
\and
Departamento de F\'isica, FCFM, Universidad de Chile, Blanco Encalada 2008, Santiago, Chile\label{aff116}
\and
Satlantis, University Science Park, Sede Bld 48940, Leioa-Bilbao, Spain\label{aff117}
\and
Instituto de Astrof\'isica e Ci\^encias do Espa\c{c}o, Faculdade de Ci\^encias, Universidade de Lisboa, Tapada da Ajuda, 1349-018 Lisboa, Portugal\label{aff118}
\and
Cosmic Dawn Center (DAWN)\label{aff119}
\and
Niels Bohr Institute, University of Copenhagen, Jagtvej 128, 2200 Copenhagen, Denmark\label{aff120}
\and
Universidad Polit\'ecnica de Cartagena, Departamento de Electr\'onica y Tecnolog\'ia de Computadoras,  Plaza del Hospital 1, 30202 Cartagena, Spain\label{aff121}
\and
Infrared Processing and Analysis Center, California Institute of Technology, Pasadena, CA 91125, USA\label{aff122}
\and
INAF, Istituto di Radioastronomia, Via Piero Gobetti 101, 40129 Bologna, Italy\label{aff123}
\and
Department of Physics, Oxford University, Keble Road, Oxford OX1 3RH, UK\label{aff124}
\and
Aurora Technology for European Space Agency (ESA), Camino bajo del Castillo, s/n, Urbanizacion Villafranca del Castillo, Villanueva de la Ca\~nada, 28692 Madrid, Spain\label{aff125}
\and
ICL, Junia, Universit\'e Catholique de Lille, LITL, 59000 Lille, France\label{aff126}
\and
ICSC - Centro Nazionale di Ricerca in High Performance Computing, Big Data e Quantum Computing, Via Magnanelli 2, Bologna, Italy\label{aff127}}    


%
%
   \abstract{
     \Euclid\/ is presently mapping the distribution of matter in the Universe in detail via the weak lensing signature of billions of distant galaxies. The weak lensing signal is most prominent around galaxy clusters, and can extend up to distances well beyond their virial radius, thus constraining their total mass.
     Near the centre of clusters, where contamination by member galaxies is an issue, the weak lensing data can be complemented with strong lensing data. Strong lensing information can also  diminish the uncertainty due to the mass-sheet degeneracy and provide high-resolution information about the distribution of matter in the centre of clusters. Here we present a joint strong and weak lensing analysis of the \Euclid Early Release Observations  of the cluster Abell 2390  at $z=0.228$. Thanks to \Euclid's wide field of view of $0.5\,\mathrm{deg}^2$, combined with its angular resolution in the visible band of $\ang{;;0.13}$ and sampling of $\ang{;;0.1}$ per pixel \cite{EuclidSkyOverview}, we constrain the density profile in a wide range of radii, 30\,kpc $< r <$ 2000\,kpc, from the inner region near the brightest cluster galaxy to beyond the virial radius of the cluster.  We find consistency with earlier X-ray results based on assumptions of hydrostatic equilibrium, thus indirectly confirming the nearly relaxed state of this cluster. We also find consistency with previous results based on weak lensing data and ground-based observations of this cluster. From the combined strong+weak lensing profile, we derive the values of the viral mass 
     $M_{200} = (1.48 \pm 0.29)\times10^{15}\, \Msun$, 
     and virial radius 
     $r_{200} =(2.05\pm0.13 \, {\rm Mpc}$), 
     with error bars representing one standard deviation. The profile is well described by an Navarro-Frenk-White model with concentration $c=6.5$ and a small-scale radius of 230 kpc in the  30\,kpc $< r <$ 2000\,kpc range that is best constrained by strong lensing and weak lensing data.  
     Abell 2390 is the first of many examples where \Euclid data will play a crucial role in providing masses for clusters. The large coverage provided by \Euclid, combined with the depth of the observations, and high angular resolution, will allow to produce similar results in hundreds of other clusters with rich already available strong lensing data.
   }
%
%
\keywords{Gravitational lensing; Galaxies: clusters: individual: Abell Clusters of Galaxies 2390; Cosmology: dark matter}
%
%

   \titlerunning{Strong and Weak lensing analysis of Abell 2390}
   \authorrunning{Diego et al.}
   
   \maketitle
%
%
%
%

\section{\label{sc:intro}Introduction}


\Euclid, a mission of the European Space Agency (ESA), is currently surveying large portions of the sky (totalling $\approx 14\,000\,\mathrm{deg}^2$ at the end of the survey) in one visible band, \IE,  and three (\YE, \JE, and \HE) near-infrared bands. A summary of the \Euclid mission is given in \cite{EuclidSkyOverview}. One of \Euclid's primary goals is to extract accurate cosmological information with its high-quality shape measurements of $\approx 1.5$ billion galaxies \citep{EP-Congedo}. These shapes contain information about massive structures along the line of sight that produce a warping of space around them, distorting the apparent shape of background galaxies. 
Many of these galaxies will be found behind galaxy clusters, allowing one to constrain the masses of galaxy clusters using gravitational lensing. \Euclid is expected to detect $\approx 10^6$ clusters with masses above $10^{14}\Msun$ and up to $z\approx 2$ \citep{EuclidSkyOverview}. All of these clusters will imprint weak lensing signatures on background galaxies observed by \Euclid and some of them will also act as strong gravitational lenses, producing multiple images of background galaxies. 

Galaxy clusters are the most powerful gravitational lenses. The combined capabilities of \Euclid regarding sensitivity of the VIS and NISP instruments \citep{EuclidSkyVIS,EuclidSkyNISP}, spatial resolution, and sky coverage, allow us to extract the best weak lensing (WL) signal around these clusters. Some of the clusters observed by \Euclid contain multiply lensed galaxies (or arcs),  which can be used as constraints on the distribution of mass in the central regions of clusters, and also to break the mass-sheet degeneracy in the inner region constrained by strong lensing, provided that the lensed galaxies are at different redshifts \citep{Saha2000,Normann2024}. 
At large radii, where only weak lensing data is available, some level of mass-sheet degeneracy may still introduce some uncertainty.
\Euclid's spectroscopic capabilities are more limited in the crowded fields of galaxy clusters, making the determination of redshifts for these strongly lensed arcs a challenging task. However, many of the most prominent galaxy clusters observed by \Euclid have already been observed by other space telescopes, including the \HST (HST), and to a lesser degree the \textit{James Webb} Space Telescope (JWST), together with ground-based telescopes that were used to estimate the redshifts of many of these strongly lensed galaxies. The ancillary strong lensing (SL) data available on these clusters (catalogues of spectroscopically confirmed multiply lensed galaxies) present a unique opportunity to combine precise SL data in the inner core of massive clusters with \Euclid's WL data extending far beyond the virial radius of each cluster. Such combinations of SL and WL data have already been performed in the past.
Earlier efforts combined space-based data for the SL constraints with ground-based data for the WL constraints \citep{Broadhurst2005,Limousin2007,Oguri2009,Sereno2011,Umetsu2016,Chiu2018,Abdelsalam1998,Bradac2006,Umetsu2010,Jauzac2012,Wong2017}, 
but in such cases the quality of the WL data was affected by seeing, which limits the accuracy of the shape measurements,  as well as the depth of the observations. In other cases, only space-based data were used, but these are restricted to small areas around the central region of the cluster, thus limitating the results \citep{Niemiec2023,Cha2024,Zitrin2015,Chiu2018,Patel2024}.
A recent compilation of methods used in the literature for joint weak and strong lensing analysis can be found in \cite{Normann2024}.

\Euclid's data are a very valuable addition to pre-existing SL data sets, since they provide the sky coverage that is lacking in earlier space-based observations of clusters, while being unaffected by the atmospheric effects plaguing ground-based WL observations. Of particular interest are clusters at relatively low redshift with known multiply lensed galaxies. Low-redshift clusters have more background galaxies behind them and cover larger areas that can be easily covered by \Euclid. 

One such low-redshift galaxy cluster lens, Abell 2390 (or A2390), was observed as part of the Performance Verification phase of \Euclid in late 2023. The targets observed during this campaign form a subset of observations known as the \Euclid Early Release Observations (ERO) which  were made public in May 2024. 
A description of the \Euclid ERO programme is given in \cite{EROData}. The preliminary results on A2390 were presented in \cite{EROLensData}, where a first analysis of the WL data was included. However, the results presented in \cite{EROLensData} do not take advantage of new photometric redshift estimates for galaxies in the \Euclid image.

\begin{figure*}
  \centering
  \includegraphics[width=18cm]{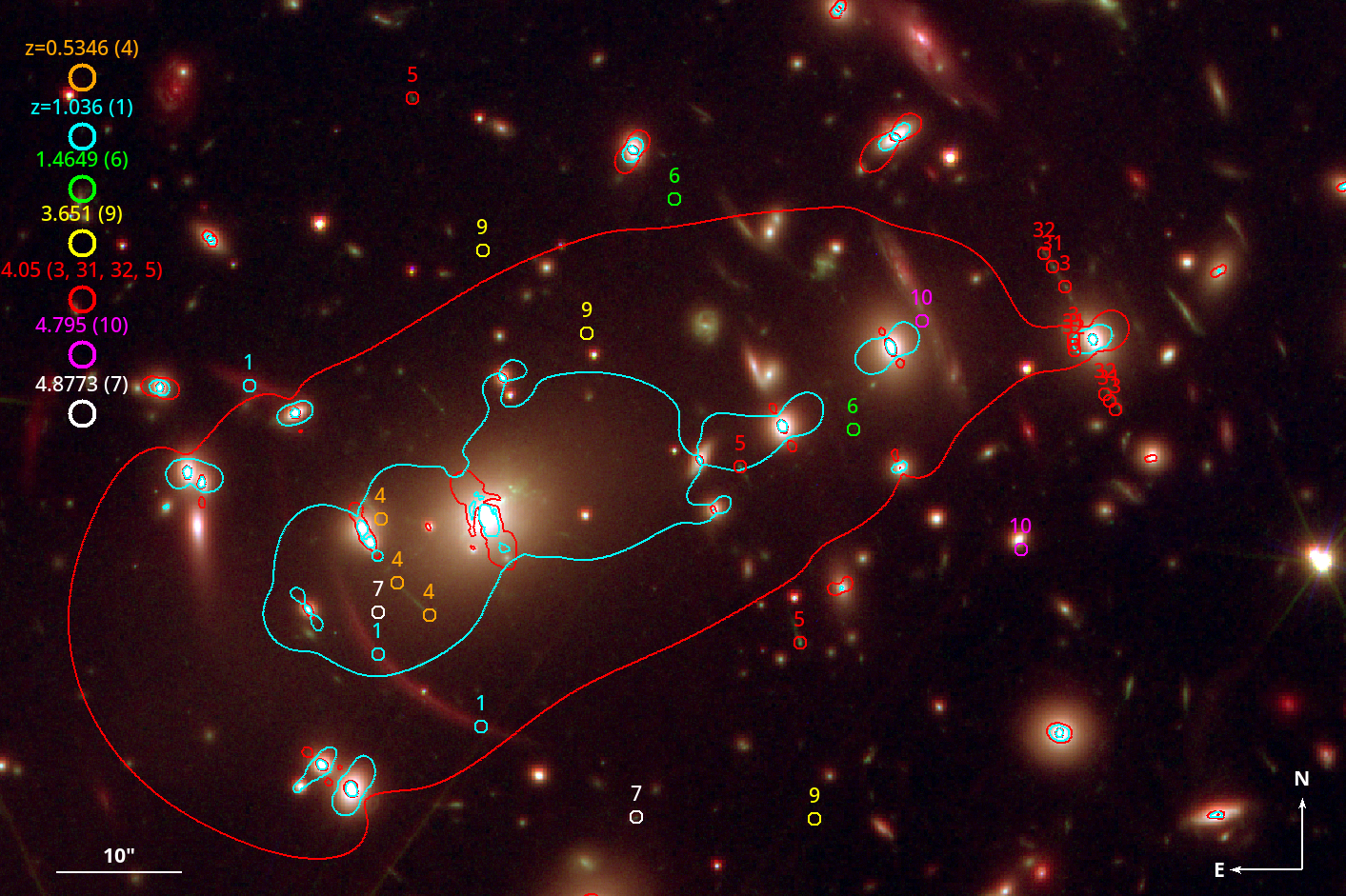}  
  \caption{Colour composite image (Red=NISP $\JE+\YE+\HE$, Green=HST-ACS-F850LP, Blue=VIS \IE) of the central region of A2390. The circles mark the position of the spectroscopic sample of SL constraints from \cite{Richard2008,Richard2021} colour coded by their spectroscopic redshift (indicated in the upper left). The number next to each small circle is the ID for each family of counterimages. 
  We show the critical curves from the SL+WL model 
  in cyan at $z_{\rm s}=1.036$ and red at $z_{\rm s}=4.05$. 
  The image covers $\ang{;1.73;}\times\ang{;1.18;}$. 
  }
  \label{fig:A2390_Arcs}
\end{figure*}  

A2390 is a massive cluster at $z = 0.228$. From earlier work, its virial mass is estimated to be  $M_{200} \approx 1.7\times 10^{15}\, \Msun$ and it has an estimated velocity dispersion of $\sigma_v \approx 1100$ km\,s$^{-1}$ \citep{Carlberg1996}. 
Lens models for A2390 were derived in the past from ground-based observations and HST images \citep{Pierre1996,Frye1998,Pello1999,Swinbank2006,Richard2008} but in all cases restricted to strong lensing.
\Chandra X-ray observations show A2390 as a cool-core cluster with a temperature $\kB T=(11.5\pm1.5)$ keV in the outskirts \citep{Allen2001}. Good agreement is found between the inferred mass profile (under the assumption of hydrostatic equilibrium) and a Navarro--Frenk--White (NFW) profile with scale radius 0.8 Mpc and concentration $c\approx 3.3$ \citep{Allen2001}.
More recently, additional evidence for a massive cooling flow towards the brightest cluster galaxy (BCG) is discussed in \cite{Alcorn2023} based on extended line emission found surrounding the BCG. Furthermore, ALMA observations reveal a massive concentration of molecular gas around the BCG, coincident with the X-ray and optical emissions \citep{Rose2024}.
Radio observations of A2390 suggest past merger activity and the presence of polarised radio emission at the edges of the cluster \citep{Bacchi2003}. Finally, high-resolution LOFAR images of A2390 show double-tailed emission around the BCG, strongly suggesting the presence of an active galactic nucleus at its centre \citep{Savini2019}.

The paper is organised as follows. In Sect.~\ref{sc:data} we describe the weak and strong lensing data used in this work.
 In Sect.~\ref{sc:WSLAP+} we discuss the algorithm \wslap \citep{Diego2005,Diego2007} used to combine the weak and strong lensing data and derive the mass distribution.
 We present our main results in Sect.~\ref{sc:results}.
 We finally conclude in Sect.~\ref{sc:conclusions}. 
We adopt a flat cosmological model with $\Omega_{\rm m}=0.3$ and $H_0=70$ km\,s$^{-1}$\,Mpc$^{-1}$. For this model, and at the redshift of the lens ($z=0.228$), $\ang{;;1}=3.65$ kpc.

\section{\label{sc:data}Euclid and ancillary data on A2390}
 Our WL measurements come entirely from the new \Euclid data. Here we summarise their main characteristics and refer the interested reader to \cite{Schrabback2025} for further details.
\Euclid observations of this cluster were carried out on November 28th 2023. The $5\sigma$ limiting magnitudes of the stacked images in the 4 \Euclid filters are \IE = 27.01 for VIS (in apertures with diameter $\ang{;;0.3}$) and \YE = 25.18, \JE = 25.22, \HE = 25.12 for the NISP images
(apertures with diameter $\ang{;;0.6}$ ).

The photometric information (and derived photometric redshifts) were obtained after combining information from \Euclid and ground-based ancillary data, in particular data from Subaru/Suprime-Cam \citep{miyazaki02} in the bands $B$, $V$, $R_c$, $i$, $I_c$, and $z'$ (covering $\approx\ang{;28;}\times\ang{;34;}$), and Canada-France-Hawaii Telescope (CFHT) Megacam u-band. Both data sets complement \Euclid observations with photometric measurements,  relevant for the estimation of  photometric redshifts. The ground-based data do not cover the full extent of the \Euclid imaging, so we restricted our main analysis to a smaller square area of $\ang{;26.85;}\times\ang{;26.85;}$ where ground-based data are available and photometric redshifts are most reliable. This area is approximately 2.5 times smaller than the entire field of view of our \Euclid data, but is perfectly centred on the galaxy cluster. 
A2390 is at low Galactic latitude ($b\approx -28^{\circ}$), 
where cirrus emission is significant. Photometry and shape measurements are performed over the 'denebulised' images, where the contribution from cirri has been reduced. The PSF was modeled with {\small PSFex} \citep{Bertin2011}, and the mask from \cite{EROLensData} is used, after expanding it with extended low-redshift galaxies. Photometric redshifts are derived from \Euclid and ground-based data using {\small Phosphoros} \citep{EP-Paltani}, specifically developed for the Euclid Science Ground Segment and based on the code {\small LePhare} \citep{Arnouts1999,Ilbert2006}. In addition, a self-organizsing map approach combining all photometric information (ground-based and space-based data), and callibrated with COSMOS data, is also used to construct the distribution of galaxies as a function of redshift, $n(z)$. See \cite{Schrabback2025} for details.

In this work we use weak lensing measurements from a subsample of background galaxies (from our photometric redshift catalogue) observed by \Euclid. We refer to this subsample as the clean catalogue that was produced by vetoing galaxies that are likely members of A2390 or foreground galaxies. The clean catalogue is built after matching and merging three independent shape measurements: \texttt{SE++} \citep{bertin22,kuemmel22}; \texttt{KSB+}  \citep{kaiser1993,kaiser1995,luppino1997,hoekstra1998,erben2001,schrabback10}; and \texttt{LensMC} \citep{EP-Congedo}. This catalogue is the basis of the main weak+strong lensing analysis. In appendix A we show results obtained from the raw catalogue where no vetoing of galaxies is applied. The raw catalogue still suffers from contamination from foreground and member galaxies and it is included here only for illustration purposes.
For the clean catalogue we find a number density of sources with reliable shear measurements between $\approx 10$ and $\approx 50$ galaxies per square arcminute. This number density is representative of the expected source density for weak lensing in \Euclid. For comparison, the  number density in the raw catalogue before removing foreground and member galaxies is approximately twice as large.

For the SL portion of the data, we relied on previous studies of this cluster with HST. We used the SL sample from \cite{Richard2008,Richard2021} that complements previously known redshifts of prominent arcs in the cluster with new spectroscopic estimates from the Multi Unit Spectroscopic Explorer (MUSE). The SL constraints are shown as coloured circles in Fig.~\ref{fig:A2390_Arcs}, where the number indicates the ID of the family of multiple images from the same background galaxy. The lensed galaxies cover a wide range in redshift between $z=0.5346$ (system 4) to $z=4.877$ (system 7). The giant arc to the east (system 1) is at $z=1.036$, and for illustration purposes we show the critical curve derived from our SL+WL model at that redshift, which nicely intersects this arc close to its middle point (as expected). 
The redshifts of each system are shown in the top left portion of the figure, with numbers in parentheses indicating the system ID with that redshift. 
Figure~\ref{fig:A2390_Arcs} also shows the critical curve at $z=4.05$ (see Sect.~\ref{sc:results}).

\section{\label{sc:WSLAP+}A joint SL$+$WL solution with \wslap}
\wslap was first described in \cite{Diego2005} and was originally developed to obtain the mass of galaxy clusters in a free-form way. That is, with no assumptions about the distribution of mass. The code was later expanded in \cite{Diego2007} to include weak lensing measurements as additional constraints.  A further development is presented in \cite{Sendra2014} where the code migrated from its native free-form nature to a hybrid type of modelling where prominent member galaxies are added into the lens model with a mass distribution that matches the observed light distribution. The total mass of the member galaxies is the only free parameter, which is adjusted as part of the optimisation process. The addition of member galaxies plays a double role. On the one hand, it adds small-scale mass fluctuations that cannot be captured by the free-form part of the model. On the other hand, member galaxies act as a regularisation factor, anchoring the solution to a stable point in the space of solutions and reducing the problem of overfitting the data \citep[see][]{Sendra2014}, which can otherwise produce spurious structures around clusters as described in \cite{Ponente2011}. The code has been validated with realistic lensing simulations \citep{Meneghetti2017},  
used to reconstruct multiple lens models including all six Hubble Frontier Fields models based on HST data \citep{Diego2016,VegaFererro2019}, and more recently to model clusters observed with the \textit{James Webb} Space Telescope \citep{Diego2023,Diego2024}.
\wslap is well suited to fit rich lensing data sets where many constraints are available, as in the case of MACS0416 where  \wslap is used to fit the largest (to date) number of strong lensing constraints (observed positions of lensed galaxies) 
in a single galaxy cluster \citep[more than 300,][]{Diego2024}. The free-form and hybrid nature of \wslap is also appropriate for combining SL and WL data, since the optimisation of the solution is carried by simultaneously fitting the SL and WL constraints. \wslap was previously used to combine SL and WL data in the well studied Abell 1689 cluster combining SL data from HST with WL data from Subaru \citep{Diego2015}.

\begin{figure*}
  \centering
   \includegraphics[width=18cm]{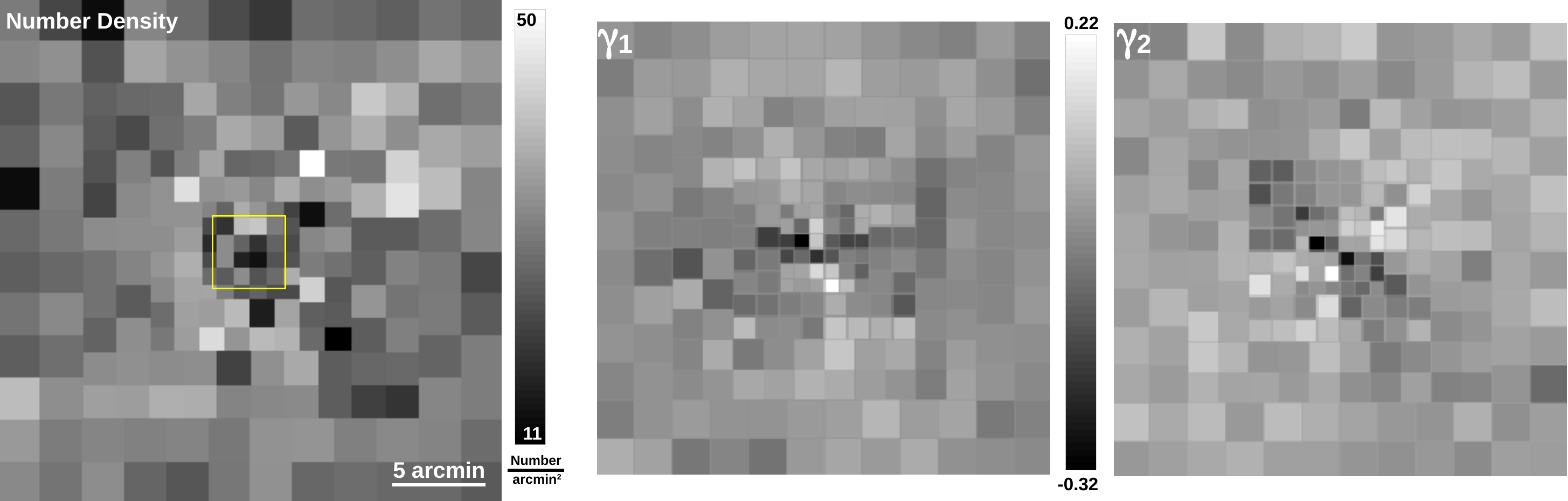}
  \caption{Multi-scale grid used for binning the shear measurements in the clean catalogue. The grid is centred in the cluster. \textit{Left}: shows the number of galaxies per square arcminute in each bin when considering only galaxies in the tomographic bins 3--10 described in \cite{Schrabback2025}.  The smallest bins in the centre are approximately $\ang{;1;}\times\ang{;1;}$ while in the edges the bins are $\ang{;2.24}\times\ang{;2.24;}$. The full area is  $\ang{;26.85;}\times\ang{;26.85;}$. A similar grid configuration is adopted for the Gaussian functions in the distribution in mass. For the mass grid, the central  $\ang{;4;}\times\ang{;4;}$ region (where the strong lensing constraints are found) is marked with a yellow square. This small region is further divided with an additional set of 123 smaller Gaussians (not shown).
  \textit{Middle and right}: show the observed shear $\gamma_1$ and $\gamma_2$ in the same multi-scale grid, same field of view, and for \texttt{LensMC}. The classic structure of $\gamma_1$ and $\gamma_2$ can be seen (i.e., horizontal and $\pi/4$-oriented quadrupole). 
  }
  \label{fig:A2390_ShearGrid}
\end{figure*}

\wslap is based on a simple decomposition of the deflection field, $\vec{\alpha}$, as a superposition of $N_{\rm G}$ two-dimensional Gaussians (other decompositions have been tested, including polynomials, but the Gaussian decomposition provides the most robust results). Under this decomposition, the deflection $\vec{\alpha}$ at a given position $\vec{\theta}$ is computed as the sum of Gaussians placed at specific predetermined positions. These positions define a grid where the Gaussians are placed. The width of each Gaussian is also determined by the grid as separation between neighbouring grid points.
\begin{equation}
\vec{\alpha}(\vect{\theta}) = \frac{4 G}{c^2} \frac{D_{\rm ds}}{D_{\rm os} D_{\rm od}} \sum_{N_{\rm G}} m_i(\vect{\theta}_i,\vect{\theta})\frac{\vect{\theta} - \vect{\theta}_i}{| \vect{\theta} - \vect{\theta}_i |^2}\; ,
\label{Eq_alpha}
\end{equation}
 where $m_i(\vect{\theta}_i,\vect{\theta})$ is the mass contained in the Gaussian mass distribution at position $\vect{\theta}_i$ and integrated out to the position $\vect{\theta}$. The factors $D_{\rm ds}$, $D_{\rm os}$, and $D_{\rm od}$ are the angular diameter distances from the deflector to the source, from the observer to the source, and the observer to the deflector, respectively. The positions of the $N_{\rm G}$ Gaussian functions, $\vect{\theta}_i$, define a grid that can be either regular in shape or follow an irregular distribution, with a higher density of points in the central region, where SL constraints are present, and the WL signal is stronger. A regular distribution of grid points represents a flat prior on the mass distribution since it makes no assumptions about the distribution of mass. The irregular grid is equivalent to a prior on the mass, where more mass is expected in regions where the grid has higher density.  For situations like the ones discussed in this paper, where the centre of the lens is well determined from SL constraints, the use of the irregular grid is well motivated.  Nevertheless, the irregular grid that we use is derived after a first optimisation performed with the regular grid, which results in an initial map of the mass distribution. The irregular grid samples this mass distribution and increases the number density of grid points in the portions of the lens plane with more mass, while reducing the total number of grid points, $N_{\rm G}$, and hence the number of free parameters. 
  The distribution of the grid is shown in Fig.~\ref{fig:A2390_ShearGrid}. The width of the Gaussians range between about $\ang{;1;}$ near the centre to about $\ang{;2;}$ near the edges. In the central $4\arcmin\times4\arcmin$ region constrained by SL (and marked with a yellow rectangle in Fig.~\ref{fig:A2390_ShearGrid}) we increase the resolution even further by adding 123 smaller Gaussians with widths ranging from  $12\arcsec$ near the BCG galaxy to $40\arcsec$ near the edges of the central $4\arcmin\times4\arcmin$ region. Combined with the 208 large-scale Gaussians that cover the larger $\ang{;26.85;}\times\ang{;26.85;}$ region, the total number of Gaussians is 331. We tested the solution with different grid configurations with varying number of grid points, $N_{\rm g}$. For solutions where $250 \lesssim N_{\rm g} \lesssim 400$ we find small variation in the solution. Lens models derived with $N_{\rm g} < 250$ start to degrade in resolution while models derived with $N_{\rm g}>400$ introduce artefacts, specially at large radii (noise overfitting).

For each SL constraint (or position  $\vec{\theta}$), we have two equations similar to Eq.~\eqref{Eq_alpha}, one for the $x$-component and one for the $y$-component of $\vect{\theta}$. Given $N_{\theta}$ lensing constraints, the lens equation ($\vect{\beta}=\vect{\theta}-\vect{\alpha}$) can be written in algebraic form 
\begin{equation}
\left(
\begin{array}{c}
\vec{\theta_x} \\
\vec{\theta_y} \\
\end{array}
\right) = \left(
\begin{array}{c}
\tens{\Theta_x} \\
\tens{\Theta_y} 
\end{array}
\right) \vec{M} + 
\left(
\begin{array}{c}
\beta_x \\
\beta_y \\
\end{array}
\right)\; ,
\label{Eq_SL_matrix}
\end{equation}
where  $\vec{\theta_x}$ and $\vec{\theta_y}$ are both an array of dimension $N_{\theta}$ containing the observed $x$ and $y$ coordinates of the $N_{\theta}$ SL arcs.  $\vec{\beta_x}$ and $\vec{\beta_x}$ are each arrays of dimension $N_{\rm s}$ containing the unknown coordinates ($x$ and $y$) of the $N_{\rm s}$ sources, and  $\vec{M}$ is another array of dimension $N_{\rm G}$ cells containing the unknown masses of the Gaussians in the grid. Finally, the matrices $\tens{\Theta_x}$ and $\tens{\Theta_y}$ each have dimension $N_{\theta} \times N_{\rm G}$ and are known once the SL constraints and grid positions are determined. The terms of this matrix are the individual elements appearing in Eq.~\eqref{Eq_alpha}, with the masses $m_i$ of each grid point set to $1\times 10^{15}\, \Msun$  at large distances from the grid points. Near each grid point, this mass is set by a Gaussian distribution,
\begin{equation}
m_i(\vec{\theta_i},\vec{\theta})=10^{15}\Msun\left(\frac{\int_0^{\vec{\theta}} G(\vec{\theta_i},\vec{\theta})\,\theta\,{\rm d}\theta}{\int_0^{\infty} G(\vec{\theta_i},\vec{\theta})\,\theta\, {\rm d}\theta}\right)\;. 
\end{equation}

Adding member galaxies to the Gaussian grid decomposition of the mass can also be expressed in same the algebraic form as Eq.~\eqref{Eq_SL_matrix}, with the only difference being that the mass of the member galaxies is a superposition of layers, with each layer containing a group of galaxies having similar light-to-mass ratio. The minimum number of layers is 1 and the maximum is the number of member galaxies considered, or lens planes in cases where line of sight effects need to be considered. 
For this work we adopt the simplest scenario where all member galaxies have the same light-to-mass ratio so there is only one layer containing all member galaxies.

The WL constraints can also be described in terms of differences of the deflection field and have a similar algebraic description to Eq.~\eqref{Eq_SL_matrix}, only in this case the left column contains the observed shear measurements ($\gamma_1$ and $\gamma_2$) and on the right-hand side there is no $\beta$ term,
\begin{equation}
\left(
\begin{array}{c}
\gamma_1 \\
\gamma_2 \\
\end{array}
\right) = \left(
\begin{array}{c}
\tens{\Gamma_1} \\
\tens{\Gamma_2} 
\end{array}
\right) \vec{M}\;.
\label{Eq_WL_matrix}
\end{equation}
Similarly, member galaxies contribute to the shear with a similar  
system of linear equations as in Eq.\;\eqref{Eq_WL_matrix}. 
Since observations measure the reduced shear, $g=\gamma/(1-\kappa)$, WSLAP+ uses a previous solution (transformed into $\kappa$) to compute the matrix elements in Eq.~\eqref{Eq_WL_matrix}. This is an important correction in regions of the lens plane where $\kappa$ is more significant (i.e, near the centre of the lens). This process is iterated twice to ensure consistency between $\kappa$ and the reduced shear.
 
The latest improvement to \wslap was the addition of the position of critical curves, based on the observed position of arcs merging at the critical curves and their orientation. We do not use critical point information in this work but we include a brief description here for completion. In  \citet[][appendix A.2]{Diego2022}  it is shown how, after a suitable rotation, the critical point information can also be described as a linear transformation of the deflection field in a form similar to the previous linear equations,
\begin{equation}
\left(
\begin{array}{c}
\vec{\lambda_1} \\
\vec{\lambda_2} \\
\end{array}
\right) = \left(
\begin{array}{c}
\tens{\Lambda_1} \\
\tens{\Lambda_2} 
\end{array}
\right) \vec{M}\;,
\label{Eq_CP_matrix}
\end{equation}
where $\lambda_1=\kappa+\gamma_1^{\rm R}$ and $\lambda_2=\gamma_2^{\rm R}$, with $\kappa$ the lens model convergence at the critical point  and $\gamma_1^{\rm R}$, $\gamma_2^{\rm R}$ the shear at the same position but after a suitable rotation in the observed direction of the arc. After this rotation  $\lambda_1=1$ and $\lambda_2=0$ at the critical points. 
The values of  $\lambda_1$ and $\lambda_2$  can be used as additional constraints, as demonstrated in \cite{Diego2022} that used these additional constraints to improve the lens model near the lensed star "Godzilla" (at $z=2.37$).

Combining all the available constraints, we can define a global system of linear equations that contains SL, WL, and critical point information (when available): 

\begin{equation}
\left( 
\begin{array}{c}
\vec{\theta_x} \\
\vec{\theta_y} \\
\vec{\lambda_1}\\
\vec{\lambda_2}\\
\vec{\gamma_1} \\
\vec{\gamma_2} \\
\end{array} \right)= \left(
\begin{array}{cccc}
\tens{\Theta_x^M} & \tens{\Theta_x^C} & \tens{I_x} &  \tens{0} \\
 \tens{\Theta_y^M} &  \tens{\Theta_y^C} &  \tens{0} &  \tens{I_y} \\
 \tens{\Lambda_x^M} &  \tens{\Lambda_x^C} &  \tens{0} &  \tens{0} \\
 \tens{\Lambda_y^M} &  \tens{\Lambda_y^C} &  \tens{0} &  \tens{0} \\
 \tens{\Gamma_1^M}   &  \tens{\Gamma_1^C}   &  \tens{0} &  \tens{0} \\
 \tens{\Gamma_2^M}   &  \tens{\Gamma_2^C}   &  \tens{0} &  \tens{0} \\
\\
\end{array}
\right) \left(
\begin{array}{c}
\vec{M} \\
\vec{C} \\
\vec{\beta_x} \\
\vec{\beta_y} \\
\end{array}
\right)\;,
\label{Eq_SLWL_matrix}
\end{equation}
where we show the explicit structure of the matrix. All terms in the $2\times2$ tensor are matrices, as described in Eqs.~(\ref{Eq_SL_matrix}), (\ref{Eq_WL_matrix}), and (\ref{Eq_CP_matrix}). We also include here the terms associated with the member galaxies, such as $\vec{C}$ on the right-hand side, which is a multiplicative factor to the fiducial mass adopted for member galaxies. The upper index M or C refers to the matrices that are multiplying the $\vec{M}$ or $\vec{C}$ terms in the vector of unknown variables.  
The $ij$ elements in the matrix $\tens{I_x}$ are 1 if the $\theta_i$ pixel contains a constraint from the $\beta_j$ source, and are 0 otherwise. The $ \tens{0}$ matrix denotes the null matrix. Equation~\eqref{Eq_SLWL_matrix} can be written in the more compact form
\begin{equation}
\vec{\Phi} =  \tens{\Gamma}\,\vec{X}\;,
\label{Eq_SLWL_matrix_simple}
\end{equation}
where $\vec{\Phi}$ is an array containing the observed positions of the arcs, critical points at different redshifts, and binned shear measurements, $ \tens{\Gamma}$ is the known non-square matrix in Eq.~\eqref{Eq_SLWL_matrix}, and $\vec{X}$ is the solution we seek; This is, a vector with all the unknowns: masses in the Gaussian decomposition ($\vec{M}$); multiplicative factors for the fiducial mass of the member galaxies ($\vec{C}$); and source positions ($\vec{\beta_x}$ and $\vec{\beta_y}$). 

Even though the number of variables is often larger than the number of constraints, a  direct inversion of the linear system of Eq.\;\eqref{Eq_SLWL_matrix_simple}, $\vec{X} =  \tens{\Gamma}^{-1}\vec{\Phi}$, is in principle possible through singular value decomposition (which also works for non-square matrices) and after ignoring the terms associated with the smallest eigenvalues. However, the solutions obtained following this method are in general quite noisy. A better approach is to define the residual array 
\begin{equation}
\vec{R} \equiv  \vec{\Phi} -  \tens{\Gamma}\,\vec{X}\;,
\label{Eq_R}
\end{equation}
and find the minimum of the scalar function $f(\vec{R}) = \vec{R}^{\rm t}\tens{\mathcal{C}}^{-1}\vec{R}$, where $\tens{\mathcal{C}}$ is the covariance matrix,  which can be approximated by a diagonal matrix with terms $\sigma_\text{SL}^{-2}$, $\sigma_\text{CP}^{-2}$, and $\sigma_\text{WL}^{-2}$, where $\sigma_\text{SL}$, $\sigma_\text{CP}$, and $\sigma_\text{WL}$ are the uncertainties in the SL positions, critical point position, and WL shear measurements. Although not thoroughly tested yet with \wslap, off-diagonal terms in the covariance matrix can also be included, thus resulting in more precise solutions, but we ignore these terms here, since they are usually much smaller than  diagonal terms. 

For this cluster, we do not use critical point positions as constraints. 
For $\sigma_\text{SL}$ we adopt an uncertainty of $\ang{;;0.2}$ or two pixels in the VIS images. For the shear measurements, we consider contributions from statistical shear measurements and from large-scale structure \citep{Jain1997,hoekstra1998}. 
That is
\begin{equation}
\sigma_\text{WL}^2 = \sigma_\text{stat}^2 + \sigma_\text{LSS}^2\;,
\end{equation}
where $\sigma_\text{stat}$ is the average ellipticity of $N_\text{shear}$ galaxies that are found in a given bin,
\begin{equation}
\sigma_\text{stat}=\frac{0.25}{\sqrt{N_\text{shear}}}\; ,
\label{Eq_sigma_stat}
\end{equation}
where we have assumed that the typical ellipticity of a galaxy before being sheared is 25\%. For $\sigma_\text{LSS}$, we follow \cite{Jain1997}, where for our cosmological model, and assuming $z_{\rm s}=1.5$, we find 
\begin{equation}
\sigma_\text{LSS}=0.0128/\Delta^{0.42}\; ,
\end{equation}
where $\Delta$ is the size of the bins (in arcmin) used to average the shear (see Fig.~\ref{fig:A2390_ShearGrid}).
For the model reconstruction, we assume a mean redshift per bin for the weak lensing measurements corresponding to the mean of $D_{\rm ds}/D_{\rm os}$ from the WL catalogue in each of the WL bins. This value fluctuates from WL bin to WL bin and it is in the range $0.6\lesssim D_{\rm ds}/D_{\rm os}\lesssim 0.7$.

By construction, the minimum of $f(\vec{R})$ is also a solution of the linear system of Eq.~\eqref{Eq_SLWL_matrix_simple}. This can be proven by simply setting the derivative of $f(\vec{R})$ to zero. The minimum of $f(\vec{R})$ can be found by fast algorithms such as the bi-conjugate gradient, but this can result in unphysical solutions where the masses in the Gaussian decomposition (or elements $X_i$ in the solution array $\vec{X}$) can be negative. Instead, we use a somewhat slower but more robust quadratic programming algorithm that seeks the minimum with the physical constraint $X_i>0$, that is, the Gaussian masses, the mass of the member galaxies, and the source, positions all have to be positive \citep[see][for details]{Diego2005}. The positive nature of the masses is obvious. For the source positions, we define the origin of coordinates in the bottom-left corner of the field of view considered. With this choice, all positions are always positive in this frame of reference, so the requirement $X_i>0$ is meaningful.  In practice, since we are making the assumption that the distribution of mass is well represented by a superposition of Gaussian functions, and we know that this assumption must be wrong to some degree, we are not interested in the minimum of $f(R)$, but in a value of $f(R)$ close enough to the minimum, but above it by some value $\epsilon$. Without the addition of member galaxies, \wslap can recover solutions with very small values of $\epsilon$. These solutions can be affected by artefacts, such as spurious rings of dark matter \citep{Ponente2011}, which try to compensate for the imperfect original assumption with an imperfect solution. Fortunately, the addition of member galaxies reduces, and in practical terms, eliminates this problem, since the optimisation reaches a point where the addition of more fluctuations in the Gaussian masses is anchored by the fixed mass distribution inside the member galaxies. This optimisation converges to an attractor solution that is stable and does not need to be regularised \citep[see][]{Sendra2014}.

\begin{figure}
  \centering
\includegraphics[width=90mm]{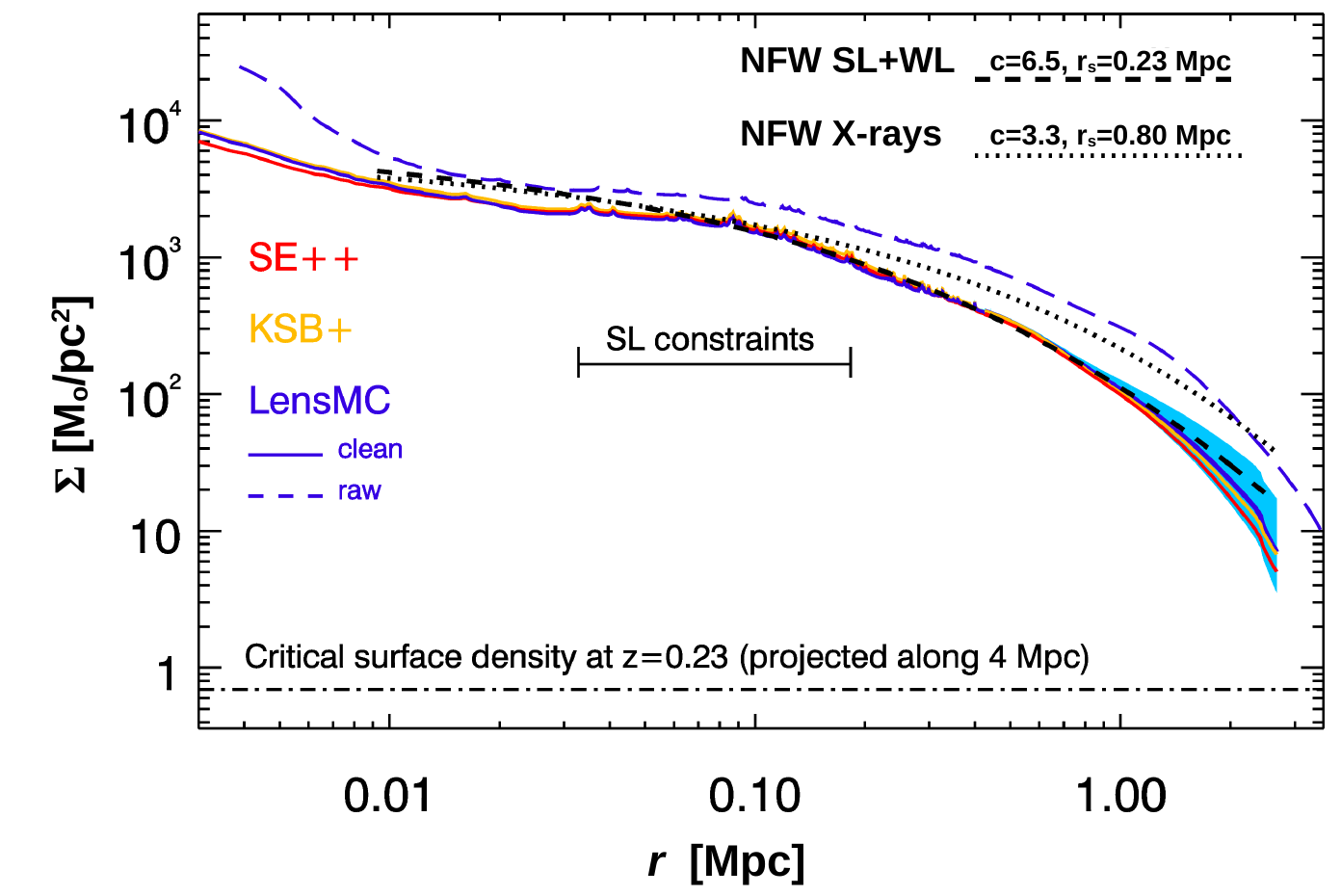}
  \caption{Surface mass density ($\Sigma$) profiles as a function of radius ($r$) centred on the BCG for the three WL clean catalogues. Two NFW profiles are shown in black-grey. The black dotted line is the NFW derived from X-rays assuming hydrostatic equilibrium. The black short-dashed line is an NFW that fits the SL+WL model in the region that is best constrained by the SL+WL data. The light-blue band shows the $1\,\sigma$ variability in the \texttt{LensMC} profile when we change the initial guess in the SL+WL optimisation algorithm.  The dark-blue dashed curve is the joint SL+WL solution when we use the raw LensMC in a larger field of view and with a minimal selection of objects (see appendix~\ref{sc:lensmc_fullcat}). 
  }
  \label{fig:A2390_profile}
\end{figure}

\section{\label{sc:results}Derived SL$+$WL lens model}
We derive a series of solutions considering the same set of SL constraints, but with the three WL catalogues described in Sect.~\ref{sc:data} (\texttt{SE++}, \texttt{LensMC}, and \texttt{KSB+}), and produced independently from the same \Euclid data \citep[see][for details]{Schrabback2025}. To explore the range of solutions that are consistent with the data, we also vary the initial guess in the optimisation process. This initial guess introduces a degree of randomness in the derived solution, which is important to explore. Regions in the lens plane playing a less significant role in fitting the data are poorly constrained and suffer from a memory effect where the value they adopt after the optimisation correlates with the random value assigned in the first step of the optimisation. This happens for instance at relatively large distances from the centre of the cluster, where the WL signal is weakest, and can be partially described by the lensing effect from the central overdensity, thus requiring little or no mass in the outskirts of the cluster. 
We explore this range of solutions for just one of the WL clean catalogues (\texttt{LensMC}). Most of the variability in the derived solution comes from the initial guess, and not the WL catalogue that is similar in all three cases as shown by \cite{Schrabback2025}. Hence, solutions derived with the other two catalogues (\texttt{SE++}, and \texttt{KSB+}) exhibit a similar degree of variability.

We optimise the mass in the lens plane with \wslap for the three WL clean catalogues (\texttt{SE++}, \texttt{LensMC}, and \texttt{KSB+}) and the fixed set of SL constraints. The mass profile obtained from each catalogue (centred in the BCG) is shown in Fig.~\ref{fig:A2390_profile}. 
The horizontal dot-dashed line in the bottom shows the critical density of the Universe at $z=0.23$ projected along 4 Mpc (roughly twice the virial radius). 
The three profiles, derived with the same initial guess, are shown as coloured lines and demonstrate excellent consistency among them, with small differences in the inner 10 kpc, which is the range unconstrained by SL measurements (or WL data). The derived profile is also in excellent agreement with the profile derived using strong lensing data only but a parametric model for the mass distribution \citep{SLonly2025}.
In Fig.~\ref{fig:A2390_profile} we also compare the derived SL+WL profiles with an NFW profile \citep{NFW1996}. This profile is given in its canonical form:
\begin{equation}
\rho(r) = \frac{\rho_{\rm o}}{r/r_{\rm s}(1+r/r_{\rm s})^2}\;,
\end{equation}
where the normalization factor, $\rho_{\rm o}$, is such that the mean density within the radius $r_{200}$ is 200 times the critical density at redshift $z$, $\rho_{\rm crit}(z)$ :
\begin{equation}
\rho_{\rm o}=\rho_{\rm crit}(z)\delta_{\rm c}=\frac{3H^2(z)}{8\pi G}\frac{200}{3}\frac{c^3}{\ln(1+c)-c/(1+c)}\;,
\end{equation}
where $\delta_{\rm c}$ is a constant that depends on the concentration parameter, $c$, and $H(z)$ the Hubble parameter at redshift $z$. We find the profile is reasonably well reproduced by an NFW with concentration parameter $c=6.5$ and a relatively small-scale radius $r_{\rm s}=230$ kpc (black dashed curve in the figure). This is higher than  the concentration derived by \cite{Allen2001} based on \Chandra data ($c\approx 3$, black dotted curve in the figure), but our best-fit NFW model agrees better with the ones derived from Subaru WL data, $c\approx 6.4$ \citep{Oguri2010} and $c\approx 6.2$ \citep{Okabe2010}. A posterior analysis in \cite{Okabe2016} finds 
$c= 4.1_{-1.0}^{+1.1}$. This lower concentration is partially explained by 
the fact that \cite{Okabe2010} used blue galaxies, which give a higher concentration (and lower masses) due to contamination of surrounding galaxies. 

The range of possible solutions obtained when we vary the initial guess is shown as a light blue band {\rm in Fig.~\ref{fig:A2390_profile}}, which can be seen only at large radii ($R\gtrsim 1$ Mpc) since at smaller radius the light-blue region is very thin. As expected, the model is less constrained at larger radii, where the WL signal is weakest. The low-end of the range of profiles is obtained when the initial guess for the optimisation is set to random normal variables with a small dispersion ($\sigma_{\rm G}=10^{11}\, \Msun$) in the initial guess for the mass of the Gaussians. The high-end of the profile range is obtained when the initial guess is allowed to have larger masses ($\sigma_{\rm G}=4\times10^{11}\, \Msun$). In both cases, Gaussians in the outskirts of the cluster play a smaller role in fitting the observed WL data (and no role in the SL portion of the data). They are poorly optimised, maintaining values closer to their original value (memory effect). 
Part of the uncertainty at large radii shown by the blue band can be attributed to the mass-sheet degeneracy which is affecting more the WL portion of the data. As the initial guess varies between small and large values of $\sigma_{\rm G}$, this is similar to applying a constant sheet of mass that retains some of its value at large radius while it is better constrained at small radii (by the SL portion of the data). This is similar to the more general ring-like degeneracy studied in \cite{Liesenborgs2008}, but see also \cite{Ponente2011} for a discussion of these artifacts.

\begin{figure}
  \centering
  \includegraphics[width=80mm]{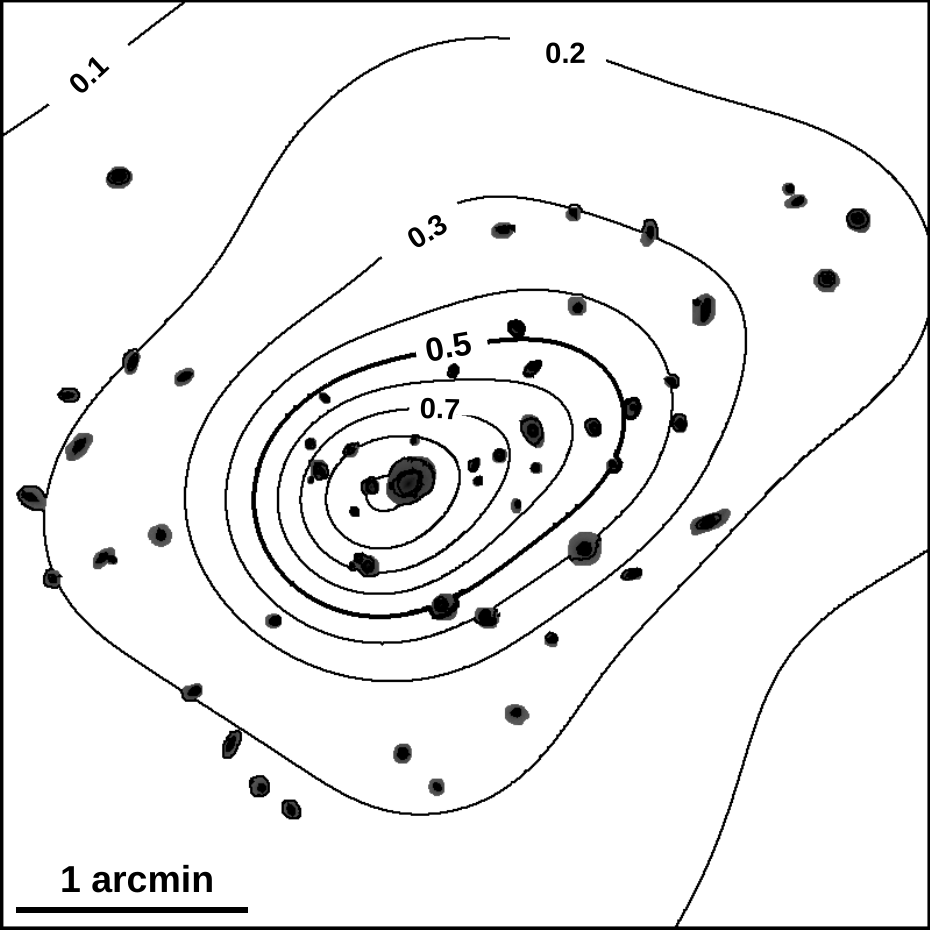}
  \caption{Contours of the convergence at $z_{\rm s}=3$. The member galaxies used in the model are also shown in black. The value of the convergence is indicated for some contours. We mark with a ticker contour the value of $\kappa=0.5$ which indicates the approximate position of the cluster critical curve at higher redshift. The value for the convergence in the last contour near the centre of the plot is $\kappa=1.06$.}
  \label{fig:A2390_Mass_Contour}
\end{figure}

From the range of possible models (light-blue shaded area), we can estimate the range of values for the virial radius, $r_{200}$. Instead of relying on an analytical fit to the observed profiles, we compute the mass density from the derived SL+WL profiles inside a cylinder of radius $R$ and length $L=2R$ along the line of sight. Then we define $r_{200}$ as the radius of the cylinder where the density inside the cylinder is 200 times the critical density. Using the cylindrical projection has the advantage of not relying on profiles to compute the overdensity in a sphere. Taking into account the dispersion of the profiles shown in Fig.\;\ref{fig:A2390_profile} (light-blue region), we estimate the virial radius to be  
 $r_{200} =(2.05\pm0.13) \, {\rm Mpc}$. 
This result is consistent with the low-end of the range of virial radii derived from X-rays in \cite{Allen2001}, $r_{200} =2.6^{+0.4}_{-0.6}$ Mpc. 
Similarly, we find that the mass enclosed within the virial radius is 
$M_{200} = (1.48 \pm 0.29)\times10^{15}\, \Msun$, 
This is consistent with previous results for this cluster, in particular the mass derived from X-rays in \cite{Allen2001}, $M_{200}=1.9^{+0.9}_{-1.0}\times10^{15}\, \Msun$, and the weak lensing masses derived from Weighing the Giants \citep[from CFHT and Subaru data,][]{Applegate2014} that found $M(<1.5\,{\rm Mpc})=1.02^{+0.18}_{-0.17}\times10^{15}\, \Msun$. Also based on Subaru data, \cite{Oguri2010} found $M_{\rm vir} = 1.04^{+0.46}_{-0.28}\times10^{15}\, \Msun$ and \cite{Okabe2010} found $M_{\rm vir} = 1.17^{+0.27}_{-0.23}\times10^{15}\, \Msun$.
More recently, \cite{Sereno2025} estimates a dynamical mass of $M_{200} = (2.63\pm 0.31)\times10^{14}\, \Msun$ for A2390, and based on the velocity dispersion of 353 members. The larger mass is partially due to the larger radius, $r_{200} =2.64\, {\rm Mpc}$, considered in that work.
The quantities $M_{\rm vir}$ and $M_{200}$ are closely related but not identical, with the ratio between the two depending on cosmology and profile parameters. $M_{\rm vir}$ is often higher than $M_{200}$ by around $20\%$ \citep{White2001}, a direct consequence of the smaller contrast in the definition of the virial mass and radius (178 versus 200 for the overdensity). The revision from \cite{Okabe2016} yields a higher mass $M_{200} = 1.51_{-0.24}^{+0.27}\times10^{15}\, \Msun$. This higher mass may be connected with the lower concentration,  $c= 4.1_{-1.0}^{+1.1}$, found in that work.  
A high mass was also found in \cite{Hoekstra2015} as part of the Canadian Cluster Comparison Project (CCCP), where they estimated $M_{\rm vir} = 2.31_{-0.36}^{+0.38}\times10^{15}\,  \Msun$ based on a fit to an NFW profile (and also for $h=0.7$). This large difference with our mass estimate cannot be attributed to the slightly different definitions of $M_{\rm vir}$ and $M_{200}$. Our mass estimate is also in excellent agreement with the WL-only result (based on the same \Euclid WL data) from \cite{Schrabback2025}, that finds  $M_{200}=(1.45 \pm 0.19)\times10^{15}, (1.64 \pm 0.21)\times10^{15}$, and $(1.49 \pm 0.21)\times10^{15}\, \Msun$ for the \texttt{SE++}, \texttt{LensMC}, and \texttt{KSB+} catalogues ,respectively, assuming an NFW halo with a concentration $c=4$.



\begin{figure}
  \centering
  \includegraphics[width=80mm]{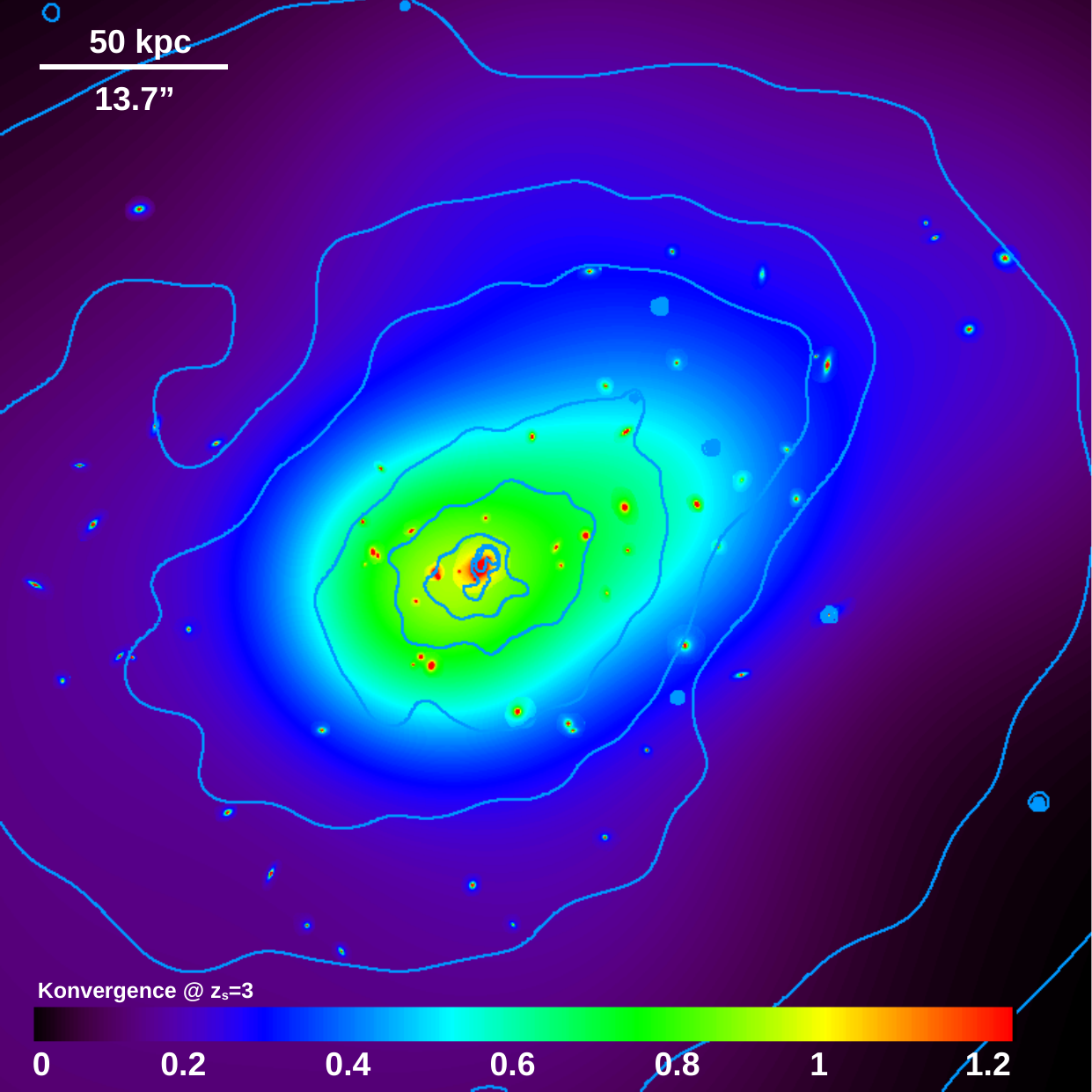}
  \caption{Comparison of the convergence map in the centre of the cluster (shown in Fig.~\ref{fig:A2390_Mass_Contour}, colour scale) with the X-ray emission from \Chandra (contours).
  For clarity, the convergence is saturated beyond $\kappa=1.2$. }
   \label{fig:A2390_Mass_Chandra}
\end{figure}

The 2-dimensional distribution of mass in the central $\ang{;4;}\times\ang{;4;}$ region is shown in Fig.~\ref{fig:A2390_Mass_Contour}. It also shows the cluster members used as part of the lens model. The contours show the values of the convergence at $z_{\rm s}=3$ and are mostly given by the smooth superposition of Gaussian functions. The smooth component peaks at the BCG position but is elongated in the NW-SE direction. A similar elongation is observed in the distribution of the X-ray emitting plasma, as shown in Fig.~\ref{fig:A2390_Mass_Chandra}. In this case, the peak of the X-ray emission aligns with the position of the BCG, as well as with the maximum in the smooth component of the mass. 
A similar result is obtained when comparing with the intracluster light (ICL) in A2390, where the peak of the ICL matches well the X-ray and mass peaks \citep{Ellien25}.
The absence of additional peaks in the smooth distribution of mass, with a relatively small elongation, suggests a relaxed dynamical state of the cluster, which may explain the good agreement between the SL+WL profile and the X-ray profile obtained after assuming hydrostatic equilibrium. On the other hand, the large concentration inferred from the SL+WL analysis favours a prolate geometry for the cluster, with its main axis offset from the line of sight and along the NW-SE direction, which could also explain the observed elongation.  
\begin{figure}
  \centering
  \includegraphics[width=90mm]{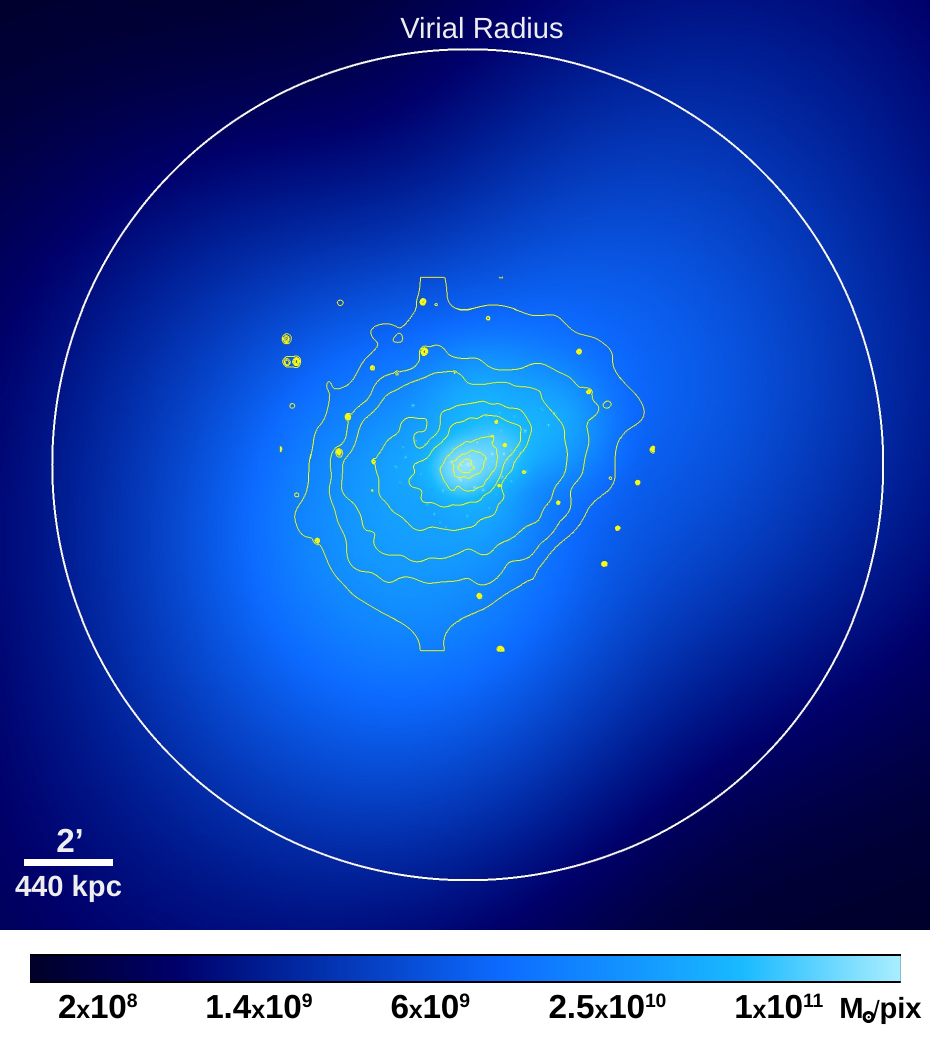}
  \caption{Mass model (colour in log scale) vs X-ray emission (contours) on a larger region and for the LensMC WL clean catalogue. Masses are given in $\Msun$ per pixel (1 pixel$=\ang{;;2}\times\ang{;;2}$ or 53.92 kpc$^2$). The  white circle has a radius of 9.36 arcmin (2.05 Mpc at the redshift of the cluster) marking the estimated virial radius of the cluster. X-ray contours correspond to 0.065, 0.12, 0.2, 0.5, 1, 2, 5, and 10 counts per pixel.}
  \label{fig:A2390_Mass_vs_Chandra_LargeFoV}
\end{figure}  

The WL data over the large field of view of \Euclid allow us to study the distribution of mass up to, and beyond, the virial radius. The two-dimensional distribution is shown in Fig.~\ref{fig:A2390_Mass_vs_Chandra_LargeFoV}. The same NW-SE elongation found in the central region is also observed on larger scales, although less well defined, and with substructure deviating from a purely elliptical shape. A slightly higher concentration of mass can be observed in the NW section of the cluster, compared with the SE sector. A small dip in the mass is also observed SW of the BCG, which appears to correlate with a reduction in the X-ray emission. However, the X-ray data are still too noisy to extract solid conclusions about this possible correlation.

\section{\label{sc:conclusions}Conclusions}

We present the first SL+WL analysis of a galaxy cluster based on \Euclid data (WL) and previous SL data. 
\Euclid will provide similar measurements of the virial mass for hundreds of clusters, with already available ancillary strong lensing high-quality data from telescopes such as HST or JWST.
The depth, large field of view, and resolution of \Euclid images allow us to detect galaxies down to $\IE \approx 27$ mag,
well beyond the virial radius of the cluster, and with spatial resolution of $\approx \ang{;;0.3}$, well suited for measurements of the shear of individual high-redshift galaxies. 

We combine the WL data from \Euclid, with preexisting HST SL data of this cluster, with the free-form hybrid method \wslap. In this approach, we model the distribution of mass as a smooth superposition of Gaussians and a more compact component that traces the light of the most prominent member galaxies in the cluster.  
For the WL data, we consider a clean catalogue, 
where we restrict ourselves to a subsample of galaxies with photometric redshifts based on \Euclid and ground-based data. 
 
From the combined SL+WL profile, we constrain the virial radius as 
  $r_{200} =(2.05\pm0.13) \, {\rm Mpc}$, 
 and virial mass as 
 $M_{200} = (1.48 \pm 0.29)\times10^{15}\, \Msun$, 
 These constraints are in agreement with earlier X-ray and weak lensing results. Additionally, the derived mass profile is consistent with earlier results based on X-ray data that assume hydrostatic equilibrium, suggesting that the cluster must be close to equilibrium. However, we find a higher concentration for the cluster, which is consistent with the higher concentration found in previous WL studies from Subaru data. We also find consistency between the derived SL+WL virial mass and virial radius and earlier results. 
 Part of the uncertainty in our results originate in the relatively small area of the ERO observations, which is further limited by the available ground-based photometry in a portion of this area. Ground-based data is required to derive photometric redshifts for all the galaxies in the field and remove both foreground and member galaxies. Future studies based on \Euclid data will cover larger areas and have improved photometric redshift estimates. This will be important to reduce the mass-sheet degeneracy that still affects the estimation of mass at large radii.

A2390 represents the first of many examples where \Euclid's high-quality WL data can be combined with available SL data to constrain the mass distribution of a cluster out to its virial radius. One of the main limitations of the current study is the lack of spectroscopic confirmation (from \Euclid) of galaxies in the same field of view, since these are not available at the time of preparing this manuscript. Future results based on \Euclid will take advantage of its spectroscopic capabilities, as well as additional ancillary observations, resulting in better removal of foreground galaxies, and improved measurements of the WL signal, and inferred virial mass. 

\begin{acknowledgements}

  \AckERO
  \AckEC
J.M.D. acknowledges support from project PID2022-138896NB-C51 (MCIU/AEI/MINECO/FEDER, UE) Ministerio de Ciencia, Investigación y Universidades.
  The scientific results reported in this article are based in part on data
  obtained from the \Chandra Data Archive (ObsId 4193, \url{https://doi.org/10.25574/04193}).  
  The analysis is derived in part using data from the Subaru telescope, the Canada-France-Hawaii Telescope, and the {\it Hubble} space telescope.
\end{acknowledgements}

%
%

\bibliography{Euclid,MyBib,EROplus}

%
%


%
%

\begin{appendix}

\section{\label{sc:lensmc_fullcat}LensMC raw shear field analysis}
\begin{figure} 
\centering
\includegraphics[angle=0,width=1\hsize]{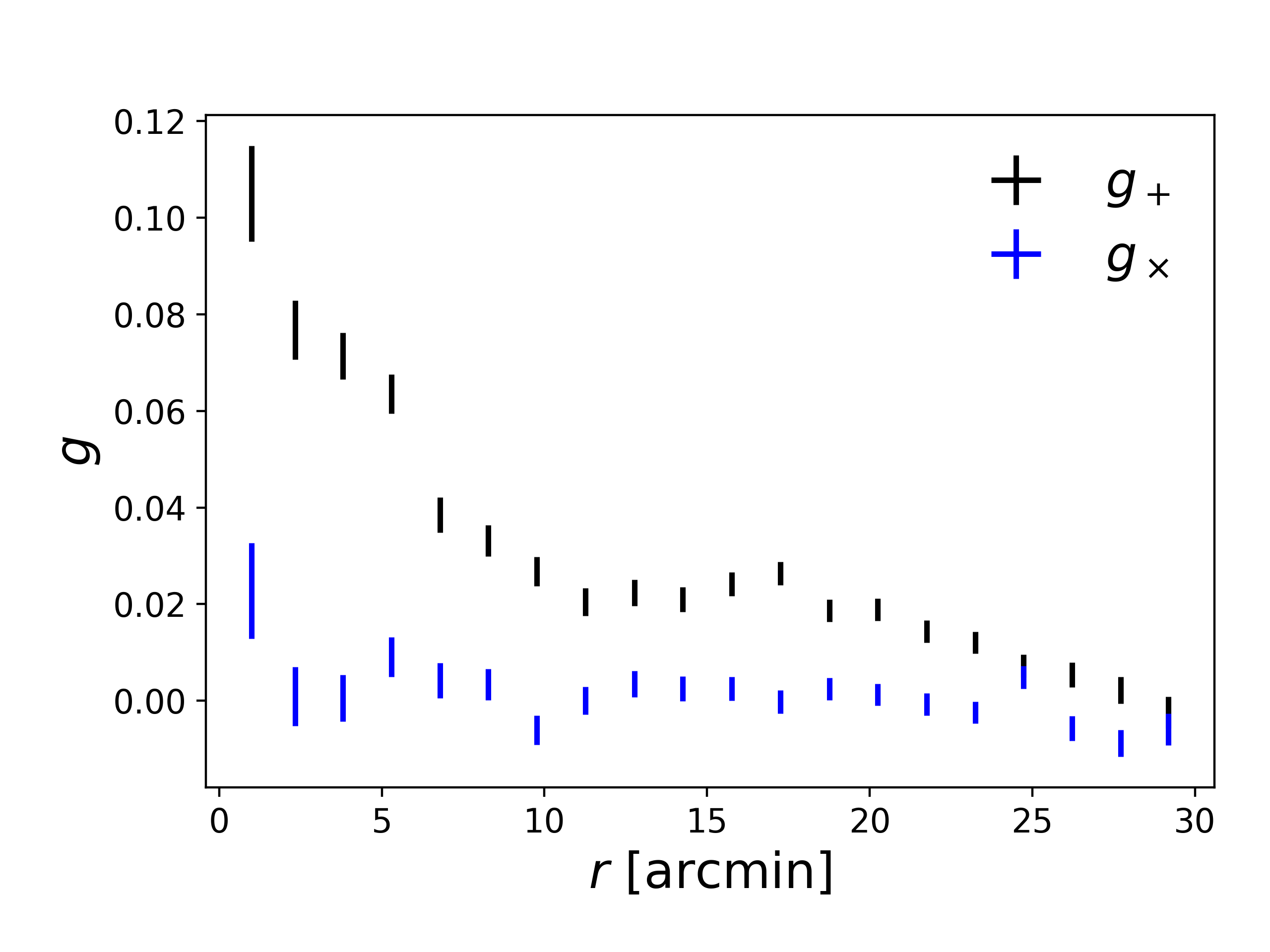}
\caption{Tangential,and cross-components,  ($g_{\rm +}$ and $g_{\rm x}$, respectively) of the raw shear estimates (without background selection and contamination correction) from the full-depth \texttt{LensMC} catalogue as a function of angular distance from the cluster centre.
The signal is significant for $r<\ang{;25;}$. The cross-shear, $g_{\rm x}$,  is largely consistent with zero (except the first bin, heavily contaminated by member galaxies), which demonstrates the good control of systematic errors in this raw catalogue.
}
\label{fig:lensmc_shear_profile}
\end{figure}

\begin{figure*} 
\centering
\includegraphics[angle=0,width=0.48\hsize]{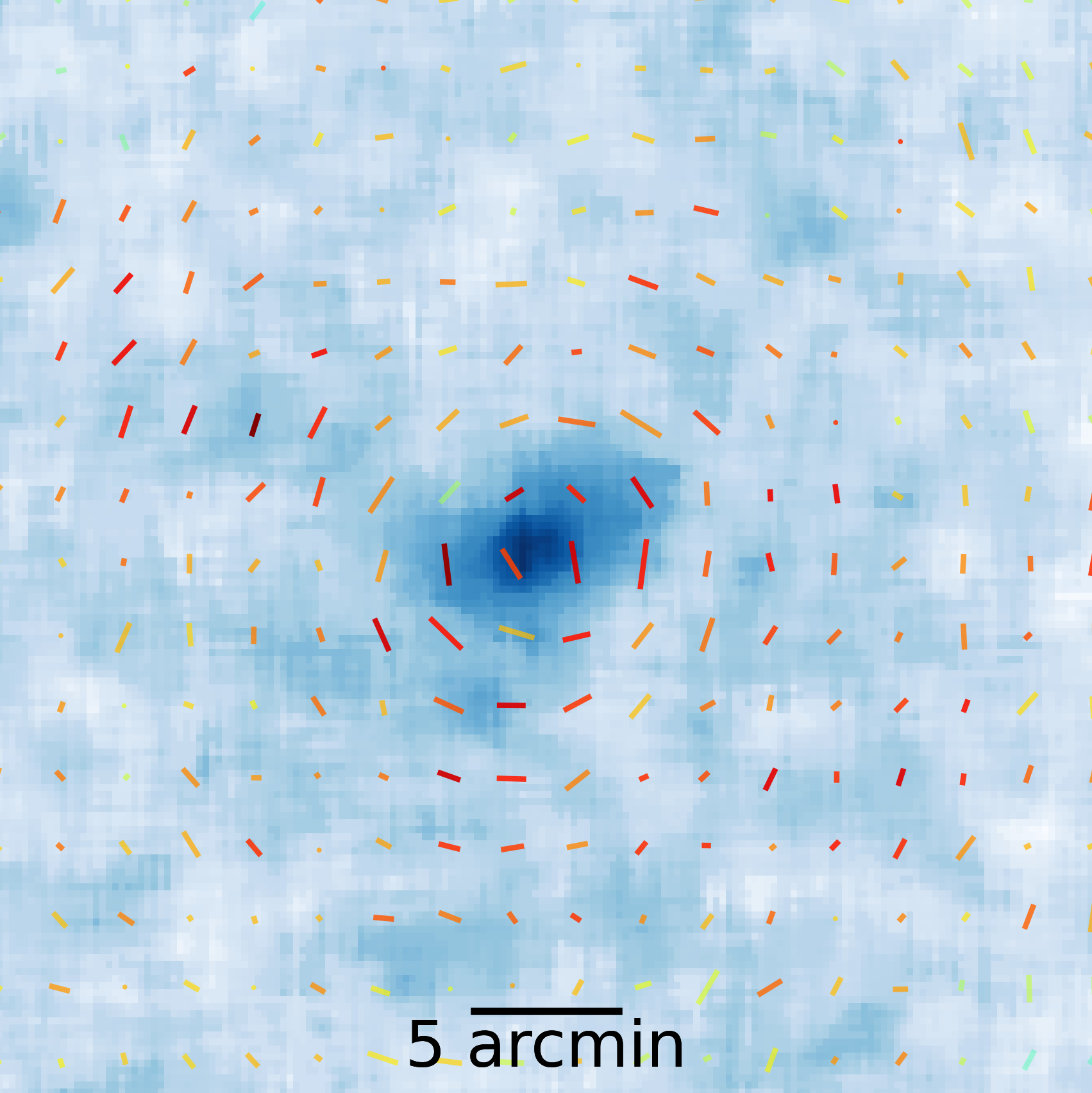}
\includegraphics[angle=0,width=0.48\hsize]{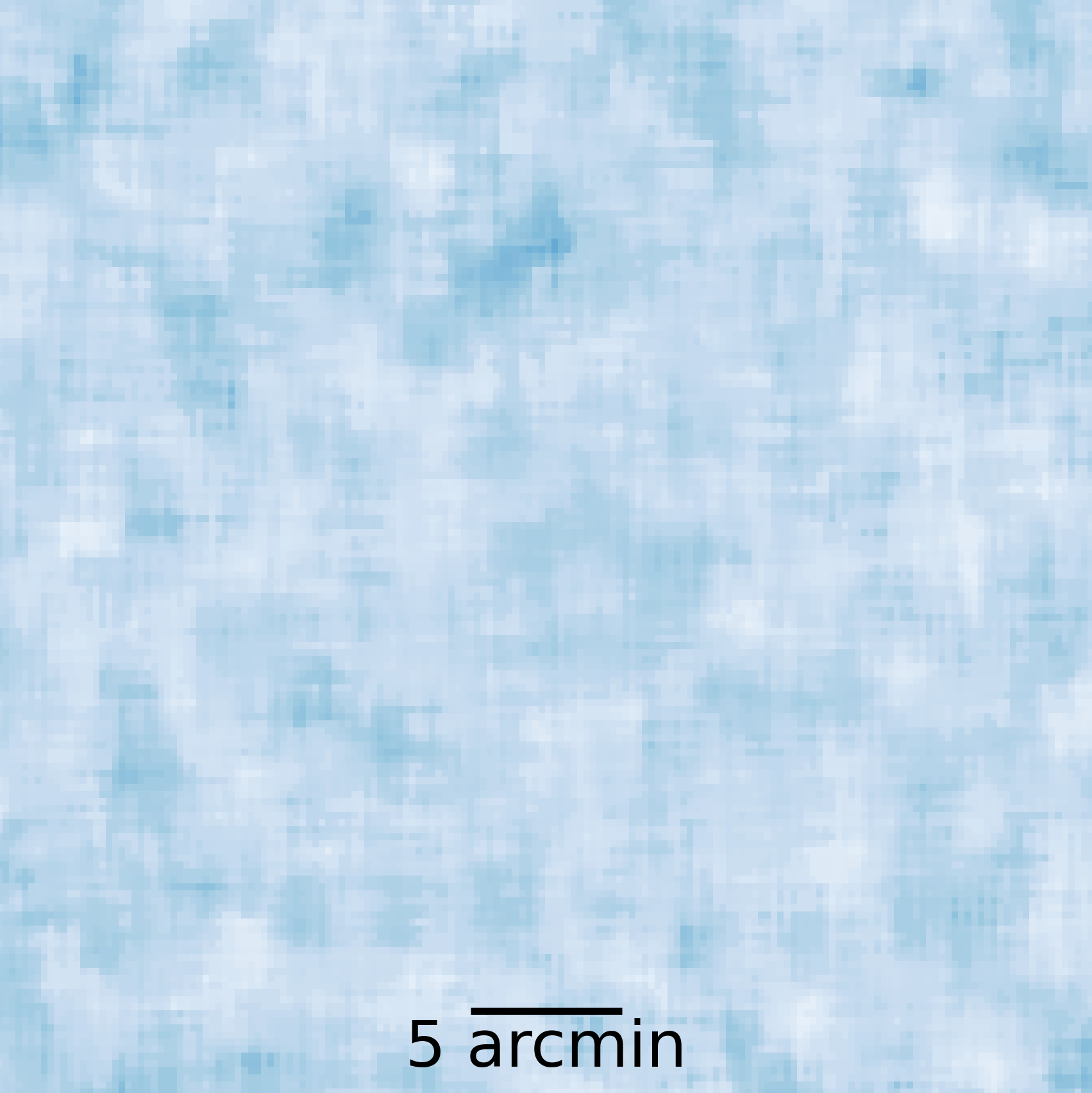}
\caption{Convergence map reconstructed from the  \texttt{LensMC} raw shear catalogue (without background selection and contamination correction), decomposed into $E$-modes (left) and $B$-modes (right). The alignment of the cluster mass in the SE-NW direction (left) is similar to the orientation found with the clean shear catalogue (Fig.~\ref{fig:A2390_Mass_Chandra}).  
The absence of a clear $B$-mode signal in the right panel demonstrates the good control of systematic errors in this raw catalogue. 
}
\label{fig:lensmc_kappa}
\end{figure*}

The main WL analysis presented in this paper is based on the clean catalogue, which is limited in sky area ($\approx0.25 \deg^2$) and depth ($\IE<26.5$) in order to facilitate the photometric selection of background galaxies and estimation of their true redshift distribution. To illustrate the full potential of the \Euclid \texttt{LensMC} shape catalogue we additionally include a simplified analysis of the measured raw shear field (without background selection, contamination correction, and redshift calibration) in this Appendix, going beyond these limitations both in terms of area and depth. Only a simple selection is done to remove unresolved objects (mostly stars) and obvious foreground objects with a larger half-light radius. This procedure removes 26.1\% of the objects in the catalogue. After this simple selection, the typical number density of sources  is larger by a factor approximately 2--3 than the number density of sources in the clean catalogue used for the main result.  The shear from this raw sample is still expected to be contaminated by member galaxies and small foreground objects. This is particularly true for the central region within the virial radius where most member galaxies are expected to be found, but as we move away from the cluster centre, member galaxies should be less of an issue in terms of contaminating the shear. In order to test the content of lensing signal in the raw shear measurements, we perform two simple tests. 

i) First, we compute the tangential and cross-components of the raw shear by weighting average of the corresponding ellipticity components in angular bins. The tangential and cross-components of the shear are computed adopting a point near the BCG as the centre,  RA$=\ang{328.406;;}$,  Dec$=\ang{17.6961;;}$.  
The  tangential component of the raw shear is detected with a signal-to-noise ratio of ${\rm S/N}=41$ when combined over all bins up to 30\,arcmin (or a factor about $3$ times the estimated virial radius).
Meanwhile, the cross component oscillates around zero, as shown in Fig.~\ref{fig:lensmc_shear_profile}.  

ii) As a second test, from the raw shear measurements we compute the convergence map via the classic Kaiser--Squires inversion \citep{kaiser1993}. The map was originally estimated on a grid size of $12\arcsec$, which is later smoothed via median filtering to $2\arcmin$. We show the resulting decomposition into $E$ and $B$ modes \citep{Stebbins1996,Hu2000,Crittenden2002,Schneider2002} in Fig.~\ref{fig:lensmc_kappa}. 
The cluster is detected with high significance in the $E$ mode reconstruction, while the $B$ mode reconstruction is consistent with noise. 
Similarly to the results obtained with the clean shear measurements, when using the raw shear measurements we find no evidence of substructure beyond the virial radius.

Finally, using the raw shear we recompute the joint SL+WL solution on the larger field covered by the raw catalogue (an area approximately twice as large as the area covered by the clean shear catalogue). The result is shown as a dashed blue line in Fig.~\ref{fig:A2390_Arcs}. The profile derived with the raw shear is clearly above the profile derived with the clean catalogue and has a smaller concentration, resembling more the profile derived from X-ray data (dotted line in the figure).  The difference with the profile obtained using the clean catalogue (solid blue line) is likely due to contamination from member galaxies and foreground objects in the raw catalogue. Also, part of the difference may be due to the fact that in the case of the raw catalogue the relative weight of the shear data (with respect to the strong lensing constraints) is larger due to the increase in the number density of shear measurements per bin (see Eq.~\ref{Eq_sigma_stat}). 

While the larger mass in the central region of the cluster may be attributed to contamination from member galaxies, the increase in mass in the dashed line profile is more noticeable at the outskirts of the cluster. Interpreting the increase in mass at larger radii is not trivial. This could still be due to contaminants to the WL signal, although we expect contamination to decrease significantly at larger radii. Alternatively, the difference may be due to the increase in the signal-to-noise ratio at larger radii, since the number density of galaxies is larger.

\end{appendix}

\end{document}